\documentclass[a4paper,twocolumn,10pt,unpublished,accepted=2024-09-02]{quantumarticle}
\pdfoutput=1
\usepackage[utf8]{inputenc}
\usepackage[english]{babel}
\usepackage[T1]{fontenc}
\usepackage{amsmath}
\usepackage{url}
\usepackage{bbold}
\usepackage{bm}
\usepackage{physics}
\usepackage{calrsfs}
\usepackage{graphicx}
\usepackage{soul}
\usepackage{bbm}
\usepackage{xcolor}
\DeclareMathAlphabet{\pazocal}{OMS}{zplm}{m}{n}
\newcommand\numberthis{\addtocounter{equation}{1}\tag{\theequation}}
\usepackage[square,comma,numbers,sort&compress]{natbib}
\usepackage{tikz}
\usetikzlibrary{shapes.geometric}
\definecolor{mygrey}{RGB}{220,220,220}
\definecolor{myblue}{RGB}{150, 173, 215}
\definecolor{mygrey}{RGB}{220,220,220}

\begin{document}

\title{Tensor-network-based variational Monte Carlo approach to the non-equilibrium steady state of open quantum systems}

\author{Dawid A. Hryniuk}
\affiliation{Department of Physics and Astronomy, University College London,
Gower Street, London, WC1E 6BT, United Kingdom}
\affiliation{London Centre for Nanotechnology, University College London, Gordon Street, London WC1H 0AH, United Kingdom}
\author{Marzena H. Szymańska}
\affiliation{Department of Physics and Astronomy, University College London,
Gower Street, London, WC1E 6BT, United Kingdom}

\maketitle

\begin{abstract}
    We introduce a novel method of efficiently simulating the non-equilibrium steady state of large many-body open quantum systems with highly non-local interactions, based on a variational Monte Carlo optimization of a matrix product operator ansatz.
    Our approach outperforms and offers several advantages over comparable algorithms, such as an improved scaling of the computational cost with respect to the bond dimension for periodic systems.
    We showcase the versatility of our approach by studying the phase diagrams and correlation functions of the dissipative quantum Ising model with collective dephasing and long-ranged power law interactions for spin chains of up to $N=100$ spins.
\end{abstract}

\section{Introduction}

The loss of quantum coherence of the system due to interactions with a dissipative environment is a fundamental limitation in the realization of many modern quantum technologies, ranging from quantum computers to quantum sensors \cite{SCHLOSSHAUER20191,RevModPhys.88.041001,RevModPhys.89.035002,RevModPhys.86.153,Harrington2022}.
While generally detrimental, in some proposals dissipation is itself harnessed to circumvent decoherence, e.g. in the dissipative preparation of entangled or topological states \cite{Budich2015,Wang2023,Roghani2018,Reiter2016,kimchi,Liu2016} or design of dissipation-stabilized states for quantum error correction 
\cite{Kempe2001, Lidar2003, PhysRevLett.85.1762, PhysRevLett.111.120501, PhysRevX.9.041053, Grimm2020, Gertler2021}.
In yet other works, the inclusion of long-ranged interactions has been proposed as a solution to some of the most pressing obstacles, e.g. by facilitating the implementation of multi-qubit codes for low-overhead fault-tolerant quantum computation \cite{multi_qubit_codes}.
Moreover, the interplay between dissipation and non-local interactions can lead to a rich variety of phenomena distinct from the coherent case \cite{PhysRevLett.130.163601,PhysRevLett.111.215305,PhysRevLett.116.113001,PhysRevResearch.2.033049,PhysRevA.97.023424}.
These theoretical investigations are being accompanied by advancements in the experimental realization of long-ranged couplings in a variety of platforms, including Rydberg atoms 
\cite{schauss,Labuhn2016,Morgado2021,Bendkowsky2009,Busche2017}, 
trapped ions 
\cite{Monroe2021,jurcevic_quasiparticle_2014,Feng2023,Pagano2020,Zhang2017}, or superconducting circuits 
\cite{PhysRevApplied.16.024018,PhysRevB.76.174507,PhysRevApplied.17.034060,PhysRevB.101.035109}. 
Despite rapid theoretical and experimental progress, the accurate characterization of driven-dissipative many-body quantum systems, particularly with highly non-local interactions, remains a challenging problem necessitating the development of new methods.

\begin{figure}[b]
    \centering
    \includegraphics[width=0.95\columnwidth]{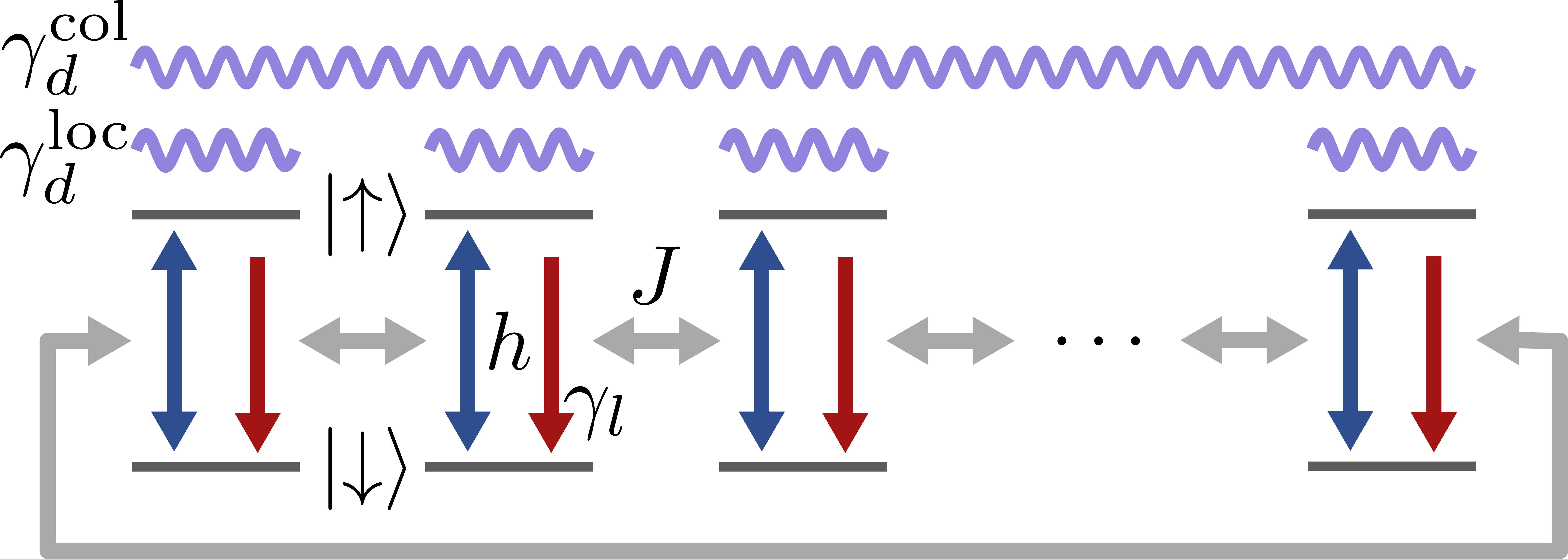}
    \caption{Schematic of an open quantum spin chain with periodic boundary conditions as an array of two-level systems. Each spin experiences coherent drive with strength $h$ and incoherent decay with strength $\gamma_l$ from coupling with the dissipative bath, and interacts with its nearest neighbors with strength $J$. Additionally, the spins may experience dephasing from contact with the dissipative bath, which may be local ($\gamma_d^\text{loc}$) or collective ($\gamma_d^\text{col}$).}
    \label{fig:Ising diagram}
\end{figure}

Open quantum systems are often well described by the Lindblad master equation. However, solving it for larger system sizes is seldom straightforward, even in one dimension, due to an exponentially increasing size of the Hilbert space (in fact, with double the exponent as compared to closed quantum systems).
Exact numerical treatment based on diagonalization of the Lindbladian superoperator or Monte Carlo averaging of stochastic quantum trajectories \cite{RevModPhys.70.101,PhysRevLett.68.580,PhysRevA.46.R6801} is possible only for small systems, while other exact approaches are limited to specific models or conditions \cite{PhysRevLett.119.190402,PhysRevA.87.042101,PhysRevLett.131.190403}.
These limitations prompted the development of many other approaches, each compromising on different aspects of the problem. 
Some popular categories include mean-field 
\cite{PhysRevB.108.054302,lee_unconventional_2013,le_boite_bose-hubbard_2014,PhysRevX.6.031011}, perturbative \cite{Li2014,PhysRevB.97.035103,PhysRevA.86.012126,PhysRevB.97.024302,huybrechts2024quantum}, phase-space \cite{PRXQuantum.2.010319,PhysRevA.105.013716,10.21468/SciPostPhys.10.2.045,PhysRevLett.128.200602}, or ensemble truncation methods
\cite{Chen2021, mccaul_fast_2021, le_bris_lowrank_2013, gravina2023adaptive}.

In the context of strongly-correlated low-dimensional quantum many-body lattices, tensor network methods, such as the celebrated density matrix renormalization group (DMRG) \cite{PhysRevLett.69.2863}, time-evolved block-decimation \cite{PhysRevLett.91.147902}, or corner transfer renormalization group in two dimensions \cite{doi:10.1143/JPSJ.65.891}, stand out as some of the most powerful approaches for closed and open systems with limited entanglement.
These methods rely on retaining only the most relevant states of the Hilbert space; the truncation is controlled by the bond dimensions of the tensor network ansatz, which can be increased to include more correlations between subsystems (at the price of costlier tensor contractions). 

For open quantum systems, most implementations rely on simulating the real-time dynamics of the system \cite{PhysRevLett.93.207204,Prosen_2009,PhysRevLett.93.207205,PhysRevLett.116.237201,Kshetrimayum_2017,PhysRevX.11.021035} (in some hybrid approaches, the dynamics is unraveled using quantum trajectories and simulated with tensor network techniques \cite{daley_quantum_2014,Jaschke_2019,PhysRevLett.128.243601,PhysRevA.110.012207,10.21468/SciPostPhys.15.4.152}), from which the steady-state properties can be extracted in the long-time limit.  
However, the potentially rapid increase in entanglement during the time evolution may cause the bond dimension to grow exponentially \cite{PhysRevLett.100.030504}, causing such methods to fail to capture the correct long-time behavior \cite{Cui, Hermes_2012}.
In this regard, when studying the steady state,
variational methods \cite{Cui,PhysRevA.92.022116,PhysRevB.105.195152,CASAGRANDE2021108060} (and related imaginary-time evolution methods \cite{PhysRevLett.119.010501})
are powerful alternatives to real-time evolution. It is known that a correct description of the steady state frequently requires a smaller bond dimension than of the intermediate states in the transient dynamics towards it \cite{PhysRevA.90.033612}.
A variational approach thus allows one to target the steady state directly, bypassing these costly intermediate states, as demonstrated in \cite{Cui}.
Nonetheless, appropriate cost functions for open quantum systems are non-linear in the Lindbladian \cite{Weimer2015}, severely limiting the effectiveness of this approach \cite{Kshetrimayum_2017}.

In recent years, variational Monte Carlo (VMC) methods based on the optimization of quantum neural networks have made strides in simulating a wide variety of open quantum
lattices in one and two dimensions \cite{Nagy2019,Vicentini2019,Hartmann,Yoshioka,Mellak2022,PhysRevLett.127.230501,PhysRevLett.128.090501,vicentini2022positivedefinite,kothe2023liouville,mellak2024deep}.
In \cite{Hartmann} it was found that a solution for the steady state of the dissipative anisotropic Heisenberg model in one dimension was possible with a neural network ansatz with close to 40 times fewer variational parameters than with a MPO-based method that achieves comparable accuracy.
However, in contrast to tensor networks, quantum neural networks are restricted to smaller system sizes, with the largest system hitherto studied in the literature with a time-independent approach limited to 20 spins \cite{Mellak2022}.
Moreover, it is unclear how well this class of variational ansätze can capture long-ranged correlations \cite{PhysRevX.11.021035}.

Inspired by previous variational tensor network and quantum neural network approaches to open quantum lattices, we formulate a new time-independent variational Monte Carlo method, based on the optimization of a MPO tensor network ansatz for the steady-state density matrix, which we term variational matrix product operator Monte Carlo (VMPOMC).
By combining the strengths of tensor networks and variational Monte Carlo, our method outperforms and enjoys distinct advantages over comparable approaches.
It is also capable of simulating computationally-challenging, highly non-local interactions, such as long-ranged Ising interactions or collective dephasing, which we investigate in this work.

This paper is organized as follows: In Sec. \ref{sec: basic concepts} we review some fundamental concepts necessary for the formulation of VMPOMC, which we then develop in Sec. \ref{sec: method}. We benchmark and illustrate the capabilities of our method in Sec. \ref{sec: results}, concentrating on variants of the paradigmatic dissipative quantum Ising model, including with long-range interactions and collective dephasing. Sec. \ref{sec: discussion} provides a summary and outlook.

\section{Basic concepts} \label{sec: basic concepts}

Our goal is to characterize the steady-state properties of one-dimensional, translationally-invariant, interacting quantum systems in contact with a dissipative environment, described by a Markovian quantum master equation in Lindblad form (setting $\hbar=1$) \cite{10.1093/acprof:oso/9780199213900.001.0001, Rivas2012},
\begin{equation} 
    \partial_t \rho = \pazocal{L}\rho = -i[H,\rho] + \sum_k \left(\Gamma_k \rho \Gamma_k^\dagger - \frac{1}{2}\{\Gamma_k^\dagger \Gamma_k,\rho\}\right),
    \label{lindblad master equation}
\end{equation}
where the Hamiltonian $H$ governs the coherent dynamics of the system and where the Lindblad jump operators $\Gamma_k$ model the incoherent dynamics due to the coupling of the system to the dissipative environment.
The non-equilibrium steady state of the system, $\rho_\text{SS}$, is defined as the solution to $\pazocal{L}\rho_\text{SS} = 0$. In this work, we only consider models with a unique steady state.

For notational convenience, we work in Liouville space \cite{Gyamfi2020, kothe2023liouville}, i.e. space of vectorized density matrices, where we map $\rho = \sum_{\{\alpha_i,\beta_i\}} \rho_{\alpha_1\dots\alpha_N}^{\beta_1\dots\beta_N} \ket{\alpha_1\dots\alpha_N}\bra{\beta_1\dots\beta_N}$ to $\ket{\rho} = \sum_{\{x_i\}} \rho_{x_1\dots x_N} \ket{x_1\dots x_N}$ by an application of Choi's isomorphism \cite{CHOI1975285,RevModPhys.93.015008} to each single-body state, abbreviating the indices by the pair $x_i = (\alpha_i,\beta_i)$, where we denote $\bm{x} = (x_1, \dots, x_N)$.
In Liouville space, superoperators are mapped to matrices; from the properties of the Kronecker product one obtains, for a generic Lindbladian in Eq. \eqref{lindblad master equation},
\begin{align*}
    \pazocal{L} &= -i( H\otimes \mathbb{1} - \mathbb{1} \otimes H^\top ) \\&\quad+ \sum_k \left(\Gamma_k \otimes \Gamma_k^* - \frac{1}{2}\Gamma_k^\dagger \Gamma_k \otimes \mathbb{1} - \frac{1}{2} \mathbb{1} \otimes \Gamma_k^\top \Gamma_k^*  \right). \numberthis \label{matricized lindbladian}
\end{align*}
When the dimensions of the Lindbladian matrix are sufficiently small, the steady state can be found by solving for the null eigenvector of \eqref{matricized lindbladian}.
One can also define the completeness relation and inner product in Liouville space. The former,
\begin{equation}
    \sum_{\bm{x}}\ket{\bm{x}}\bra{\bm{x}} = \mathbb{1}, \label{completeness relation}
\end{equation}
follows directly from the usual completeness relation. The inner product between vectorized operators $\ket{A}$ and $\ket{B}$, which we denote 
by the braket $\bra{A}\ket{B}$, is to be understood as the Hilbert-Schmidt product,
\begin{equation}
    \bra{A}\ket{B} = \tr A^\dagger B.
\end{equation}
It follows that the Hilbert-Schmidt (Frobenius) norm of the operator (matrix) $A$ can be expressed as \cite{Horn1985}
\begin{equation}
    \|A\|^2 = \bra{A}\ket{A} = \tr A^\dagger A = \sum_{\mu,\nu} |A_{\mu\nu}|^2. \label{hilbert schmidt norm}
\end{equation}

\begin{figure} 
    \scalebox{1.1}{
    \begin{tikzpicture}[square/.style={regular polygon,regular polygon sides=4, inner sep=0, minimum size=30pt}]
    
        \node[text width=3cm] at (0.5,4) {(a)};
    
        \node[] (left) at (-1,3) {};
        \node[] (leftup) at (-1,3.6) {};
        \node (B1) at (0,3) [square,draw,fill=myblue,rounded corners=3,thick] {$A$};
        \node (B2) at (1.5,3) [square,draw,fill=myblue,rounded corners=3,thick] {$A$};
        \node (B3) at (3,3) [square,draw,fill=myblue,rounded corners=3,thick] {$A$};
        \node[] (Bm1) at (3.8,3) {};
        \node[] (Bm2) at (4.2,3) {};
        \node[] (Bm1up) at (3.8,3.5) {};
        \node[] (Bm2up) at (4.2,3.5) {};
        \node[] (dots) at (4,3) {$\dots$};
        \node[] (dotsup) at (4,3.5) {$\dots$};
        \node (BN) at (5,3) [square,draw,fill=myblue,rounded corners=3,thick] {$A$};
        \node[] (right) at (6,3) {};
        \node[] (rightup) at (6,3.5) {};
    
        \draw [thick] (B1) -- (B2);
        \draw [thick] (B2) -- (B3);
        \draw [thick] (B3) -- (Bm1);
        \draw [thick] (Bm2) -- (BN);
    
        \node[] (B1b) at (0,2) {$\alpha_1$};
        \node[] (B2b) at (1.5,2) {$\alpha_2$};
        \node[] (B3b) at (3,2) {$\alpha_3$};
        \node[] (BNb) at (5,2) {$\alpha_N$};
    
        \draw [thick] (B1) -- (B1b);
        \draw [thick] (B2) -- (B2b);
        \draw [thick] (B3) -- (B3b);
        \draw [thick] (BN) -- (BNb);
    
        \node[] (B1t) at (0,4) {$\beta_1$};
        \node[] (B2t) at (1.5,4) {$\beta_2$};
        \node[] (B3t) at (3,4) {$\beta_3$};
        \node[] (BNt) at (5,4) {$\beta_N$};
    
        \draw [thick] (B1) -- (B1t);
        \draw [thick] (B2) -- (B2t);
        \draw [thick] (B3) -- (B3t);
        \draw [thick] (BN) -- (BNt);
    
        \draw [thick,rounded corners] (B1) -- (-0.6,3) -- (-0.6,3.5) -- (Bm1up);
        \draw [thick,rounded corners] (Bm2up) -- (5.6,3.5) -- (5.6,3) -- (BN);

        \node[text width=3cm] at (3.1,1.4) {vec};
        \node[] (vec1) at (2.3,2) {};
        \node[] (vec2) at (2.3,0.8) {};
        \draw [-latex,line width=2pt] (vec1) -- (vec2);

        \node[text width=3cm] at (0.5,1) {(b)};
    
        \node[] (left) at (-1,0) {};
        \node[] (leftup) at (-1,0.6) {};
        \node (A1) at (0,0) [square,draw,fill=myblue,rounded corners=3,thick] {$A$};
        \node (A2) at (1.5,0) [square,draw,fill=myblue,rounded corners=3,thick] {$A$};
        \node (A3) at (3,0) [square,draw,fill=myblue,rounded corners=3,thick] {$A$};
        \node[] (m1) at (3.8,0) {};
        \node[] (m2) at (4.2,0) {};
        \node[] (m1up) at (3.8,0.5) {};
        \node[] (m2up) at (4.2,0.5) {};
        \node[] (dots) at (4,0) {$\dots$};
        \node[] (dotsup) at (4,0.5) {$\dots$};
        \node (AN) at (5,0) [square,draw,fill=myblue,rounded corners=3,thick] {$A$};
        \node[] (right) at (6,0) {};
        \node[] (rightup) at (6,0.5) {};
    
        \draw [thick] (A1) -- (A2);
        \draw [thick] (A2) -- (A3);
        \draw [thick] (A3) -- (m1);
        \draw [thick] (m2) -- (AN);
    
        \node[] (A1b) at (0,-0.9) {};
        \node[] (A1b2) at (0.2,-0.9) {};
        \node[] (A1bm) at (0.15,-1) {$x_1$};
        \node[] (A2b) at (1.5,-0.9) {};
        \node[] (A2b2) at (1.7,-0.9) {};
        \node[] (A2bm) at (1.65,-1) {$x_2$};
        \node[] (A3b) at (3,-0.9) {};
        \node[] (A3b2) at (3.2,-0.9) {};
        \node[] (A3bm) at (3.15,-1) {$x_3$};
        \node[] (ANb) at (5,-0.9) {};
        \node[] (ANb2) at (5.2,-0.9) {};
        \node[] (ANbm) at (5.15,-1) {$x_N$};
    
        \draw [thick] (A1) -- (A1b);
        \draw [thick] (A2) -- (A2b);
        \draw [thick] (A3) -- (A3b);
        \draw [thick] (AN) -- (ANb);
    
        \node[] (A1t) at (0,1) {};
        \node[] (A2t) at (1.5,1) {};
        \node[] (A3t) at (3,1) {};
        \node[] (ANt) at (5,1) {};
    
        \draw [thick,rounded corners=1mm] (A1) -- (0,0.8) -- (0.2, 0.8) -- (A1b2) ;
        \draw [thick,rounded corners=1mm] (A2) -- (1.5,0.8) -- (1.7, 0.8) -- (A2b2) ;
        \draw [thick,rounded corners=1mm] (A3) -- (3.0,0.8) -- (3.2, 0.8) -- (A3b2) ;
        \draw [thick,rounded corners=1mm] (AN) -- (5.0,0.8) -- (5.2, 0.8) -- (ANb2) ;
    
        \draw [thick,rounded corners] (A1) -- (-0.6,0) -- (-0.6,0.5) -- (m1up);
        \draw [thick,rounded corners] (m2up) -- (5.6,0.5) -- (5.6,0) -- (AN);
    \end{tikzpicture}    
    }
    \caption{(a) The translationally-invariant MPO  with periodic boundary conditions in the matrix element of $\rho$ in  Eq. \eqref{MPO}, expressed in Penrose graphical notation. Lines connecting two nodes imply a tensor contraction. 
    (b) The MPS in the coefficient of $\ket{\rho}$ in Eq. \eqref{MPO vectorized} after vectorization of the MPO, where we merged the on-site physical indices as $x_j=(\alpha_j,\beta_j)$.} 
    \label{fig: MPO}
\end{figure}
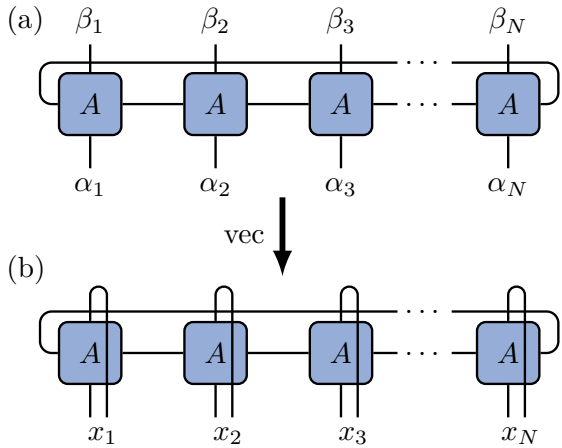

A \textit{matrix product operator} (MPO) representation of the density matrix has the form 
\begin{equation}
\rho = \sum_{\{\alpha_i,\beta_i\}} \tr A^{\alpha_1\beta_1}_{1} \dots A^{\alpha_N\beta_N}_{N}\ket{\alpha_1\dots\alpha_N}\bra{\beta_1\dots\beta_N},
\label{MPO}
\end{equation} 
where the $d\times d\times\chi\times\chi$ tensors $\{A_i^{\alpha_i\beta_i}\}_{i=1}^N$ (where $d$ denotes the local Hilbert space dimension) can be considered matrices of (bond) dimension $\chi$ for fixed physical indices $\alpha_i,\beta_i$. For translationally-invariant systems with a unit cell of size one and periodic boundary conditions, the matrices $A^{\alpha_i,\beta_i}$ are functions of the values of $\alpha_i$ and $\beta_i$ but are otherwise independent of the site $i$. In vectorized notation, after reshaping the tensors, the MPO becomes a matrix product state (MPS), 
\begin{equation}
    \ket{\rho} = \sum_{\{x_i\}} \tr A^{x_1}_1 \dots A^{x_N}_N  \ket{x_1\dots x_N}.
    \label{MPO vectorized}
\end{equation}
We depict this in Fig. \ref{fig: MPO}.

\section{Method} \label{sec: method}

Having reviewed the required prerequisites, we can now formulate the VMPOMC method. In Sec. \ref{sec: Variational Monte Carlo with Lindbladians} we set up the variational parametrization of the density matrix, introduce the notions of a cost function and the local estimator of the Lindbladian, and discuss their computation with an MPO ansatz. This is followed by a discussion of optimization of the ansatz via stochastic gradient descent and stochastic reconfiguration in Sec. \ref{sec: Optimization}.
A summary and comparison with other MPO-based approaches is provided in Sec. \ref{sec: summary}.

\subsection{Variational Monte Carlo with Lindbladians and MPOs} \label{sec: Variational Monte Carlo with Lindbladians}

In our variational approach, we represent the density matrix as a translationally-invariant MPO tensor network, $\rho=\rho(\bm{a})$, where the variational parameters $\bm{a} = \{a_i\}_{i=1}^{N_\text{param}}$ are the $N_\text{param} = d^2\chi^2$ distinct elements of the site-independent matrices $\{A^{s}\}_{s=1}^{d^2}$. For notational brevity, we use the compound index $i = (u,v,s)$ of the full set of indices labelling a tensor element of the MPO, $a_i = a_{uv}^s$. We seek to find the optimal parameter values yielding the desired steady state. To this end, we must define a suitable \textit{cost function} which attains a global minimum at the steady-state condition, and which can be systematically minimized by tuning the variational parameters to approach the steady-state condition sufficiently closely \cite{RevModPhys.93.015008,Vicentini2019}.
We resort to the variational principle, which for open quantum systems can be stated as 
$0 \leq \frac{\norm{\dot{\rho}(\bm{a})}^2}{\norm{\rho(\bm{a})}^2}$,
with the equality saturating at the steady state  \cite{Weimer2015,Vicentini2019}.
Different interpretations of the above principle are possible, depending on the chosen definition of the norm. It was argued \cite{Weimer2015, PhysRevA.91.063401} that the most optimal choice is the trace norm. 
However, the trace norm is difficult to compute in practice; instead, we shall make use of the Hilbert-Schmidt norm, defining the cost function
\begin{equation} \label{cost function}
    \pazocal{C}(\bm{a}) = \frac{\norm{\dot{\rho}(\bm{a})}^2}{\norm{\rho(\bm{a})}^2} = \frac{1}{Z}\bra{\rho(\bm{a})}\pazocal{L}^\dagger\pazocal{L}\ket{\rho(\bm{a})}, 
\end{equation}
with normalization constant (purity) $Z=\norm{\rho(\bm{a})}^2=\sum_{\bm{x}} |\langle \bm{x}|\rho(\bm{a}) \rangle|^2$. For notational brevity, we shall suppress all functional dependencies on $\bm{a}$ from now on. 

The cost function in Eq. \eqref{cost function} cannot be evaluated exactly for larger lattices as written due to exponential growth of the size of the Hilbert space.
We therefore resort to Monte Carlo sampling to evaluate it approximately. 
First, we must represent it in a form amenable to stochastic sampling. 
In addition, the product $\pazocal{L}^\dagger\pazocal{L}$ is in general difficult to handle as it's far less sparse than $\pazocal{L}$. A particularly useful form of \eqref{cost function} (originally due to \cite{Vicentini2019})---an expectation value of products of terms linear in $\pazocal{L}$---is obtained upon insertion of the completeness relation in Eq. \eqref{completeness relation} in between the Lindbladian superoperators,
\begin{align*}
    \pazocal{C}
    =& \frac{1}{Z}\sum_{\bm{x}} |\langle \bm{x}|\rho \rangle|^2 \frac{\bra{\bm{x}}\pazocal{L}\ket{\rho}^*}{\bra{\bm{x}}\ket{\rho}^*} \frac{\bra{\bm{x}}\pazocal{L}\ket{\rho}}{\bra{\bm{x}}\ket{\rho}} \\
    =& \sum_{\bm{x}} p(\bm{x}) |\pazocal{L}_\text{loc}(\bm{x})|^2 \\
    =& \mathbbm{E}_{\bm{x}\sim p(\bm{x})}[ |\pazocal{L}_\text{loc}(\bm{x})|^2 ], \numberthis \label{cost function ensemble}
\end{align*}
where we defined the \textit{local estimator of the Lindbladian}, 
\begin{equation} \label{local estimator of lindbladian}
    \pazocal{L}_\text{loc}(\bm{x}) = \frac{\bra{\bm{x}}\pazocal{L}\ket{\rho}}{\bra{\bm{x}}\ket{\rho}} = \sum_{\bm{y}} \bra{\bm{x}}\pazocal{L}\ket{\bm{y}} \frac{\bra{\bm{y}}\ket{\rho}}{\bra{\bm{x}}\ket{\rho}}.
\end{equation}
The expectation value $\mathbbm{E}_{\bm{x}\sim p(\bm{x})}[\,]$ in Eq. \eqref{cost function ensemble} can be referred to as a quantum ensemble average, in which an observable (here, $|\pazocal{L}_\text{loc}(\bm{x})|^2$) is averaged over an ensemble of many-body configurations $\{\bm{x}\}$, distributed according to the Born-like probability density $p(\bm{x}) = |\langle \bm{x}|\rho \rangle|^2/Z$.  As in classical Monte Carlo simulations, this ensemble average can be stochastically approximated by sampling a set of states $\{\bm{x}\}$ from the distribution $p(\bm{x})$. In our implementation, we draw new Monte Carlo samples via a sequential Metropolis update, updating each site of the system in fixed order, which minimizes costly tensor contractions; Monte Carlo sampling is delineated in Appendix \ref{sec: Monte Carlo sampling}.

Let us now consider more closely the expression of the local estimator of the Lindbladian. At first glance, its computation seems like a formidable task due to the sum over all basis states $\{\bm{y}\}$. In practice, however, the Lindbladian is to a large degree local, such that the subset of configurations $\{\bm{y}\}$ for which $\bra{\bm{x}}\pazocal{L}\ket{\rho}$ is nonzero is small (typically of $O(N)$).
On the other hand, because the density matrix is represented by a MPO, the brakets in the local estimator of the Lindbladian are products of matrices,
\begin{equation}
    \bra{\bm{y}}\ket{\rho} = \bra{y_1, \dots, y_N}\ket{\rho} = \tr A_{y_1}\dots A_{y_N},
\end{equation}
which are generally expensive to compute, with a formal computational complexity of $O(N\chi^3)$ for the product above. If the size of the subset of important configurations $\{\bm{y}\}$  is of  $O(N)$, this results in an overall scaling of $O(N^2\chi^3)$ per sample $\bm{x}$. However, a more careful analysis (see Appendix \ref{appendix Computation of the local estimator of the Lindbladian}) shows that, for sufficiently local Lindbladians, the above matrix products can be evaluated in sweeps where parts of previously computed matrix products 
are reused, reducing the formal computational scaling by a factor of $N$, to $O(N\chi^3)$ per sample $\bm{x}$.

It must be noted that the variational steady state can generally only be an approximation to the true steady state, owing to the finite number of variational parameters. Moreover, to constitute a physically-valid density matrix, it must satisfy: (1) hermiticity, (2) unit trace, and (3) positive semi-definiteness. While the first two conditions are easily enforced by renormalizing the matrix elements, the last is difficult to impose. Although it is possible to construct a positive definite ansatz via purification \cite{PhysRevLett.93.207204,PhysRevLett.116.237201}, the required bond dimension may become arbitrarily large \cite{Cuevas_2013}, and translationally-invariant MPOs need not admit a purified form valid for all system sizes \cite{10.1063/1.4954983}.
Therefore, we only consider the MPO in its unpurified form, which we nevertheless observe to reliably converge towards the correct, positive definite manifold of solutions.
As argued in \cite{Cui}, convergence may be formally decided by considering the variation of $\pazocal{C}$ (and other observables) as a function of increasing bond dimension, requiring that the variation falls below some set threshold value. In what follows, we shall deem "converged" simulations that attain $\pazocal{C}/N<10^{-4}$.

\subsection{Optimization} \label{sec: Optimization}

Having defined an appropriate ansatz and cost function, we now seek a systematic method of optimizing the variational parameters. We shall outline two approaches: stochastic gradient descent (SGD) and the more sophisticated stochastic reconfiguration (SR) method.

\begin{figure*}[t]
    \centering
    \includegraphics[width=0.8\textwidth]{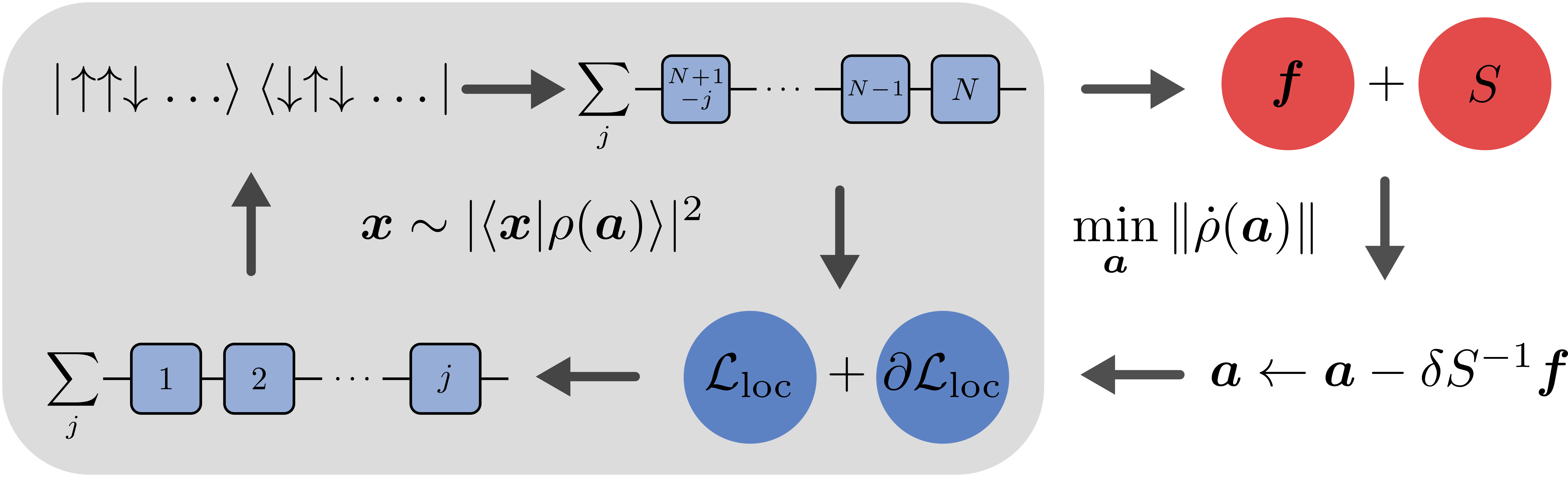}
    \caption{Graphical summary of the VMPOMC algorithm. (Inside gray box, clockwise) \textit{Sampling loop}: Given a set of variational parameters $\boldsymbol{a}$, we draw a new sample $\bm{x}$ via a leftward sequential Metropolis update, in the process of which we generate a set of right matrix products. We reuse these when computing the local estimator and local gradient tensor via a rightward sweep through the chain, during which we generate a corresponding set of left matrix products. These are then reused when drawing the next sample $\bm{x}$. (Outside gray box, clockwise) \textit{Optimization loop}: After a sufficient number of Monte Carlo samples have been gathered, the computed quantities are used to find the gradient vector and metric tensor, which are subsequently used to update the variational parameters $\boldsymbol{a}$. We iterate this process until convergence to the steady state.}
    \label{fig:method_schematic}
\end{figure*}

\subsubsection{Stochastic gradient descent}

In the gradient descent method the variational parameters are updated in the direction of the steepest decrease of the cost function $\pazocal{C}$. Although our variational ansatz is holomorphic, 
the cost function is not due to the presence of complex-conjugates of the variational parameters. 
Under these conditions, the correct expression for the steepest descent gradient vector is given by the "conjugate cogradient" \cite{kreutzdelgado2009complex}, i.e. the Wirtinger derivative of the cost function with respect to the conjugated variational parameters, $f_i= \partial_{a_i^*}\pazocal{C}$.
Let us define the diagonal superoperator
\begin{equation}
    \Delta_i = \sum_{\bm{x}} \partial_{a_i} \ln{\bra{\bm{x}}\ket{\rho}} \ket{\bm{x}}\bra{\bm{x}}
\end{equation}
satisfying
$\Delta_i \ket{\rho} = \partial_{a_i} \ket{\rho}$. In addition, we shall denote its diagonal matrix elements by $\Delta_{i}(\bm{x}) = \bra{\bm{x}}\Delta_{i}\ket{\bm{x}} = \partial_{a_i} \ln{\bra{\bm{x}}\ket{\rho}}$. 
Because $\partial_{z^*}f(z) = (\partial_{z}f(z)^*)^*$ in Wirtinger calculus we also note that $\partial_{a_i^*} \ln{\bra{\bm{x}}\ket{\rho}}^* = ( \partial_{a_i} \ln{\bra{\bm{x}}\ket{\rho}} )^* = \Delta^\dagger_{i}(\bm{x})$.
Defining the log-derivatives of the Lindbladian,
\begin{equation} \label{local derivative of lindbladian}
    \partial\pazocal{L}_{\text{loc},i}(\bm{x}) = \frac{\bra{\bm{x}}\pazocal{L}\Delta_i\ket{\rho}}{\bra{\bm{x}}\ket{\rho}} 
    = \sum_y \bra{\bm{x}}\pazocal{L}\ket{\bm{y}} \Delta_i(\bm{y}) \frac{\bra{\bm{y}}\ket{\rho}}{\bra{\bm{x}}\ket{\rho}},
\end{equation}
the expression for the gradient can be written as \cite{Vicentini2019}
\begin{equation} \label{SGD gradient}
\begin{aligned}
    f_i &= \mathbbm{E}_{\bm{x}\sim p(\bm{x})}[ \pazocal{L}_\text{loc}(\bm{x}) \partial\pazocal{L}^\dagger_{\text{loc},i}(\bm{x}) ] \\&\qquad- \mathbbm{E}_{\bm{x}\sim p(\bm{x})}[ \Delta_i^\dagger (\bm{x}) ] \times \mathbbm{E}_{\bm{x}\sim p(\bm{x})}[ |\pazocal{L}_\text{loc}(\bm{x})|^2 ].
\end{aligned}
\end{equation}
As in the case of the cost function, the ensemble averages in Eq. \eqref{SGD gradient} can be  approximated by Monte Carlo sampling, in which case we speak of stochastic gradient descent (SGD).
Likewise, the computation of the gradient can be made more manageable by exploiting the structure of the MPO ansatz;
details of the efficient computation of the gradient for an MPO ansatz are given in Appendix \ref{appendix Computation of the gradient}.

With the gradient vector known, the variational parameters can be updated by setting
\begin{equation}
    \bm{a}\leftarrow \bm{a}-\delta \bm{f}
\end{equation}
for some small step size $\delta$. 
After updating the tensor elements, the trace of the density matrix will generally be no longer unity. To alleviate this, we renormalize the local tensor elements so that $\tr\rho = 1$ after every iteration.

\subsubsection{Stochastic reconfiguration}

The above gradient descent method can be refined by accounting for the curvature of the variational manifold. Roughly speaking, small changes of the variational parameters need not correspond to small changes in the cost function; by computing the \textit{metric tensor} $S_{ij}$, it is possible to reconfigure the gradient vector $f_i$ so that the Hilbert-Schmidt distance between successive variational density matrices is minimized, accelerating convergence. This approach is known as stochastic reconfiguration (SR) \cite{becca_sorella_2017, PhysRevResearch.2.023232}. 
In the context of open quantum systems, the SR method has been adapted to the optimization of quantum neural network ansätze in \cite{Vicentini2019, Nagy2019, Yoshioka, Hartmann}. 
The resultant equations of motion for the variational parameters are 
$
    \sum_j S_{ij}\delta a_{j} = f_i 
$
\cite{becca_sorella_2017}, where the elements of the $N_\text{param}\cross N_\text{param}$ metric tensor $S_{ij}$ can be expressed as \cite{Vicentini2019}
\begin{equation}
    \begin{aligned}
        S_{ij} &= \mathbbm{E}_{\bm{x}\sim p(\bm{x})}[  \Delta_i^\dagger (\bm{x}) \Delta_j (\bm{x}) ] \\&\qquad - \mathbbm{E}_{\bm{x}\sim p(\bm{x})}[  \Delta_i^\dagger (\bm{x}) ] \times \mathbbm{E}_{\bm{x}\sim p(\bm{x})}[ \Delta_j (\bm{x}) ].
    \end{aligned}
\end{equation}
As before, we have interpreted the above expectation values as ensemble averages over the many-body configurations $\{ \bm{x} \}$, which we can evaluate stochastically, reusing the same set of samples used to calculate the cost function \eqref{cost function ensemble} and gradients \eqref{SGD gradient}.

With the metric tensor and gradient vector known, we may update the variational parameters by setting 
\begin{equation} \label{SR gradient}
    \bm{a}\leftarrow \bm{a}-\delta S^{-1} \bm{f},
\end{equation}
for some small step size $\delta$.
The above requires the inversion of the metric tensor; due to stochastic sampling and numerical inaccuracy, it may happen that $S$ becomes noninvertible. This can be prevented by adding a constant, diagonal shift to the final metric tensor,
\begin{equation} \label{SR renormalization}
    S_{ij} \leftarrow S_{ij} + \epsilon \delta_{ij},
\end{equation}
where $\delta_{ij}$ denotes the Kronecker delta and where the parameter $\epsilon$ can be tuned to modulate the strength of the reconfiguration of the gradients in Eq. \eqref{SR gradient}.

\subsubsection{Selecting optimal hyperparameter values}

The success in the search for the global minimum depends critically on the optimization hyperparameters, such as the step size for the gradient update, the number of Monte Carlo samples per iteration, or the renormalization offset when using SR.
We select appropriate values for the step size and offset by trial and error, and decrease the former geometrically with the number of iterations.
To minimize statistical inaccuracy during optimization, we ensure that a sufficiently high number of Monte Carlo samples $(N_\text{MC}\gg N_\text{params})$ \cite{becca_sorella_2017} is drawn every iteration.
In some simulations we periodically increase $N_\text{MC}$ to allow for more precise, less noise-affected updates.
Regarding the renormalization of the metric tensor, in principle it is possible to determine the minimum offset $\epsilon$ that guarantees positive semi-definiteness of the metric tensor $S$ by diagonalizing $S$ and setting the magnitude of $\epsilon$ to that of the smallest (negative) eigenvalue of $S$. However, in practice, we find that a somewhat larger offset results in a more stable convergence.

\subsection{Summary} \label{sec: summary}

For the convenience of the reader, here we outline the critical steps of a single iteration of the VMPOMC algorithm. A graphical summary is also shown in Fig. \ref{fig:method_schematic}.
\begin{enumerate}
    \item Draw a new sample from the distribution $p(\bm{x})=|\langle \bm{x} | \rho \rangle|^2/\sum_{\bm{x}} |\langle \bm{x} | \rho \rangle|^2$ via a sequential Metropolis update, while generating a set of right matrix products $\{R_j\}_{j=1}^N$.
    \item Calculate the local estimator of the Lindbladian $\pazocal{L}_\text{loc}(\bm{x}) = \mel{\bm{x}}{\pazocal{L}}{\rho}/\braket*{\bm{x}}{\rho}$, while generating a set of left matrix products $\{L_j\}_{j=1}^N$.
    \item Calculate the log-derivative tensor $\Delta_i(\bm{x})$ and local log-derivatives of the Lindbladian $\partial\pazocal{L}_{\text{loc},i}(\bm{x}) = \mel{\bm{x}}{\pazocal{L}\Delta_i}{\rho}/\braket*{\bm{x}}{\rho}$ for all parameters $i = 1, \dots, N_\text{param}$.
    \item Repeat steps (1)-(3) $N_\text{MC}$ number of times, then compute the gradients $f_i = \partial_{a^*_i} \pazocal{C}$.
    \item Optionally: compute and regularize the metric tensor $S_{ij}$.
    \item Update the tensor elements, $a_i\leftarrow a_i-\delta f_i$ \\(or $a_i\leftarrow a_i-\delta \sum_j (S^{-1})_{ij} f_j$).
\end{enumerate}

For generic $n$-local Lindbladians, the formal worst-case computational scaling for one iteration of VMPOMC with SR is $O(N_\text{MC}N^2d^{2n}\chi^3 + N_\text{MC}d^{4n}\chi^4 + d^{6n}\chi^6)$. In practice, however, the computational cost can be greatly reduced by exploiting the sparsity and diagonality of the local Lindbladian operators (see Appendix \ref{appendix Computation of the local estimator of the Lindbladian}).

\begin{figure}[t]
    \scalebox{1.1}{
    \begin{tikzpicture}[square/.style={regular polygon,regular polygon sides=4, inner sep=0, minimum size=30pt}]
        \node[] (left) at (-1,0) {};
        \node[] (leftup) at (-1,0.6) {};
        \node (A1) at (0,0) [square,draw,fill=myblue,rounded corners=3,thick] {$A$};
        \node (O1) at (0,-1) [square,draw,fill=mygrey,rounded corners=3,thick] {$\pazocal{O}_1$};
        \node (A2) at (1.5,0) [square,draw,fill=myblue,rounded corners=3,thick] {$A$};
        \node (A3) at (3,0) [square,draw,fill=myblue,rounded corners=3,thick] {$A$};
        \node[] (m1) at (3.8,0) {};
        \node[] (m2) at (4.2,0) {};
        \node[] (m1up) at (3.8,0.5) {};
        \node[] (m2up) at (4.2,0.5) {};
        \node[] (dots) at (4,0) {$\dots$};
        \node[] (dotsup) at (4,0.5) {$\dots$};
        \node (AN) at (5,0) [square,draw,fill=myblue,rounded corners=3,thick] {$A$};
        \node[] (right) at (6,0) {};
        \node[] (rightup) at (6,0.5) {};
    
        \draw [thick] (A1) -- (A2);
        \draw [thick] (A2) -- (A3);
        \draw [thick] (A3) -- (m1);
        \draw [thick] (m2) -- (AN);
    
        \node[] (A1b) at (0,-0.9) {};
    
        \node[] (A3b) at (3,-0.9) {};
    
        \draw [thick] (A1) -- (O1);
    
        \node[] (A1t) at (0,1) {};
        \node[] (A2t) at (1.5,1) {};
        \node[] (A3t) at (3,1) {};
        \node[] (ANt) at (5,1) {};
    
        \draw [thick,rounded corners=2mm] (A1) -- (0,0.8) -- (0.5, 0.8) -- (0.5,-1.8) -- (0,-1.8) -- (O1) ;
        \draw [thick,rounded corners=1mm] (A2) -- (1.5,0.8) -- (1.7, 0.8) -- (1.7,-0.8) -- (1.5,-0.8) -- (A2) ;
        \draw [thick,rounded corners=1mm] (A3) -- (3.0,0.8) -- (3.2, 0.8) -- (3.2,-0.8) -- (3.0,-0.8) -- (A3) ;
        \draw [thick,rounded corners=1mm] (AN) -- (5.0,0.8) -- (5.2, 0.8) -- (5.2,-0.8) -- (5.0,-0.8) -- (AN) ;
    
        \draw [thick,rounded corners] (A1) -- (-0.6,0) -- (-0.6,0.5) -- (m1up);
        \draw [thick,rounded corners] (m2up) -- (5.6,0.5) -- (5.6,0) -- (AN);
    \end{tikzpicture}
    }
    \caption{The computation of $\langle \pazocal{O}_1 \rangle = \tr(\pazocal{O}_1\rho)$ with $\rho$ in MPO form. Here we assumed that $\pazocal{O}_1$ is a one-body operator.}
    \label{fig: observable computation}
\end{figure}
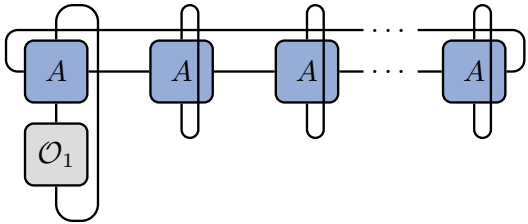

Compared to other variational MPO-based approaches \cite{Cui, PhysRevA.92.022116, CASAGRANDE2021108060}, our method enjoys numerous distinct advantages. It disposes of the need to find an MPO representation for the Lindbladian, whose bond dimension may be large and significantly impact the difficulty of the required tensor contractions \cite{PhysRevA.92.022116}.
It is readily applied to finite periodic systems; comparable MPO- or DMRG-based algorithms would have effective computational costs of $O(N^2\chi^5)$ or $O(N^2\chi^6)$ per iteration \cite{Sandvik2007,SCHOLLWOCK201196}.
It relies on the evaluation of single-layer contractions ($\bra{\bm{x}}\ket{\rho}$), 
whose computational cost is far lower than of the multi-layer contractions in a deterministic approach, further compounded by the added complexity of handling the non-local squared Lindbladian $\pazocal{L}^\dagger\pazocal{L}$ (partly alleviated in our method thanks to Eq. \eqref{cost function ensemble}). 
Due to its inherent stochasticity, it has a reduced likelihood of becoming trapped in a local minimum, and therefore does not require additional warm-up protocols as in e.g. \cite{Cui, CASAGRANDE2021108060}. 
Moreover, it is particularly suitable for treating purely- or mostly-diagonal superoperators (contingent on sufficient expressibility of the MPO ansatz) due to cancellation of the probability amplitudes in the expressions for the cost function and gradient vector, enabling exact treatment of otherwise prohibitively difficult non-local interactions, as we shall demonstrate presently.

\begin{figure}[t]
    \includegraphics[width=\columnwidth]{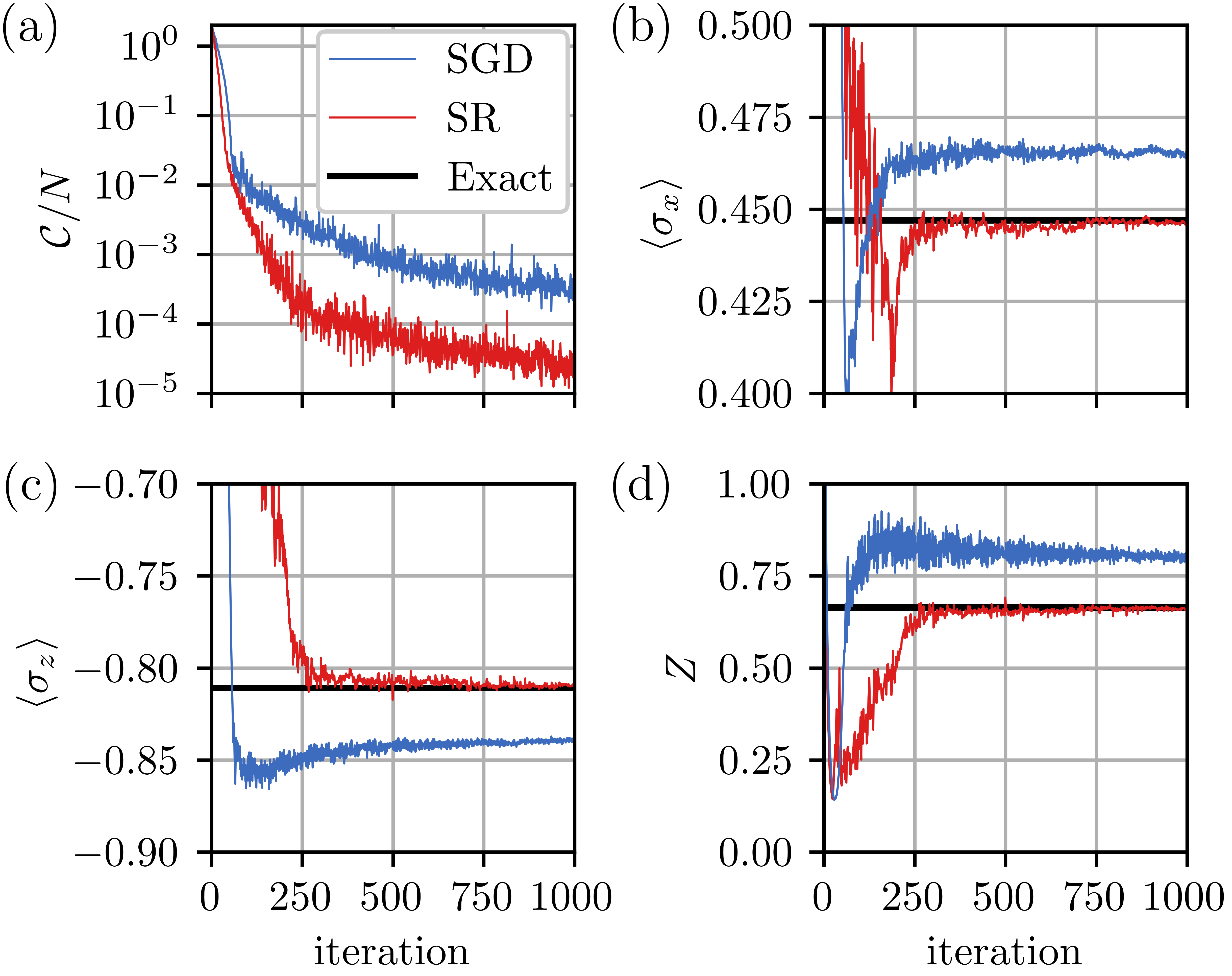}
    \caption{Sample optimization process with SGD (blue) and SR (red, $\epsilon=0.01$) over 1000 iterations for the DQIM with $N=12$ at $h=0.5$, starting from the same random initial MPO. We plot (a) the evolution of the cost-function, (b) the steady-state $x$- and (c) $z$-magnetizations, as well as (d) the purity of the variational steady-state. Shown via black horizontal lines in plots (b)-(d) are the true values of the corresponding observables obtained via an exact diagonalization of the Lindbladian. The recorded wall times for the 1000 iterations and associated measurements on an Intel i7 12700k machine were 65s for SGD and 88s for SR.}
    \label{fig:SGD SR comparison}
\end{figure}

\section{Numerical results} \label{sec: results}

We benchmark and illustrate the capabilities of VMPOMC by studying variants of the paradigmatic dissipative quantum Ising model (DQIM).
Given the variational steady-state density matrix $\rho$, we measure the steady-state expectation value of an observable $\pazocal{O}$ by computing 
$
    \langle \pazocal{O} \rangle = \tr(\pazocal{O}\rho)
$.
We note that, unlike in comparable neural-network-based algorithms \cite{Vicentini2019,Nagy2019,Hartmann}, here the expectation values can be computed exactly, i.e. without carrying a Monte Carlo error, via an appropriate tensor contraction (see Fig. \ref{fig: observable computation}). The parameter and hyperparameter values used in all subsequent simulations are detailed in the table in Appendix \ref{app: Parameters and hyperparameters}.

\begin{figure}[t]
    \includegraphics[width=\columnwidth]{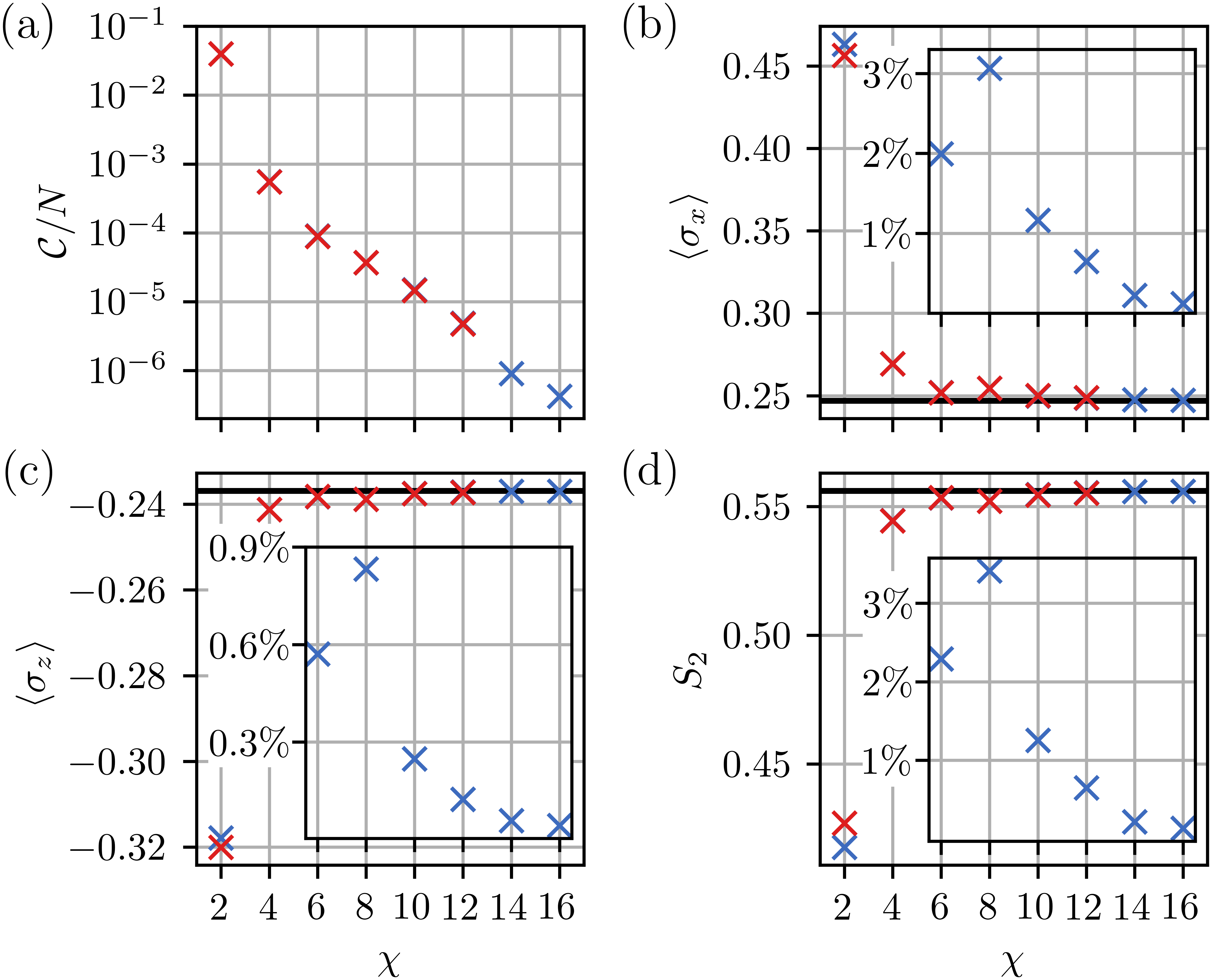}
    \caption{Convergence of VMPOMC for the DQIM with $N=12$ (blue crosses) and $N=100$ (red crosses) spins at $h=1.0$ as a function of the bond dimension $\chi$, showing the final recorded values of (a) the cost function, (b) steady-state $x$- and (c) $z$-magnetizations, and (d) steady-state Rényi-2 entropy. 
    Black horizontal lines in (b)-(d) indicate corresponding exact steady-state expectation values for $N=12$. Insets: relative errors between the variational and exact results for $N=12$ and $\chi\geq 6$ as a function of $\chi$. }
    \label{fig:Bond dimension comparison}
\end{figure}

\subsection{Dissipative Ising chain} \label{sec: dissipative ising chain}

As a first example we study the one-dimensional DQIM in a transverse magnetic field with nearest-neighbor interactions and incoherent decay, defined by the Hamiltonian
\begin{equation}
    H = J\sum_i \sigma^z_i \sigma^z_{i+1} + h\sum_i \sigma^x_i \label{dissipative transverse field ising model}
\end{equation}
with the Lindblad spin decay operator
$
    \Gamma_k = \frac{\sqrt{\gamma_-}}{2}(\sigma_k^x-i\sigma_k^y),
    \label{spin decay jump operator}
$
for all $k=1,\dots,N$, where $\sigma^n$ for $n\in(x,y,z)$ denote Pauli spin operators.
A schematic of the model (with added dephasing) is shown in Fig. \ref{fig:Ising diagram}.
Non-trivial steady-state properties of the model arise from the competition between unitary and dissipative dynamics. Variations of the DQIM have been used as a testing ground for other methods in earlier publications \cite{Vicentini2019,Yoshioka,Cui,Hartmann,PhysRevX.11.021035}. Moreover, after inclusion of long-ranged interactions, the DQIM becomes experimentally realizable with e.g. Rydberg gases \cite{PhysRevA.90.011603,PhysRevA.91.063401,PhysRevA.99.043404}. We assume $\gamma_-=2J=1$ in all subsequent simulations.

In Fig. \ref{fig:SGD SR comparison} we depict an example optimization process, comparing the SGD and SR algorithms. We plot the stochastic estimates of the cost function and a selection of steady-state observables as a function of the iteration step for the DQIM with $N=12$ sites and an MPO ansatz of bond dimension $\chi=4$. The small number of sites allows us to gauge the accuracy of our method by comparing our results against corresponding exact steady-state expectation values, which we find via an exact diagonalization of the Lindbladian. The exact results are indicated on the plots with black, horizontal lines. SR is found to converge to the minimum much more rapidly than SGD, yielding more accurate estimates of the measured observables at only a modest increase in the computational overhead. Moreover, observable estimates from SGD plateau some distance away from the exact values, possibly becoming trapped in a local minimum due to the low number of variational parameters.
It should be noted that the effectiveness of SR over SGD depends strongly on the model and chosen parameters. Nevertheless, it is found by the authors to be the superior choice in virtually all circumstances if adequately regularized. As such, we shall resort to SR exclusively in all subsequent computations (with suitably adapted values of $\epsilon$). 

\begin{figure}[t]
    \includegraphics[width=\columnwidth]{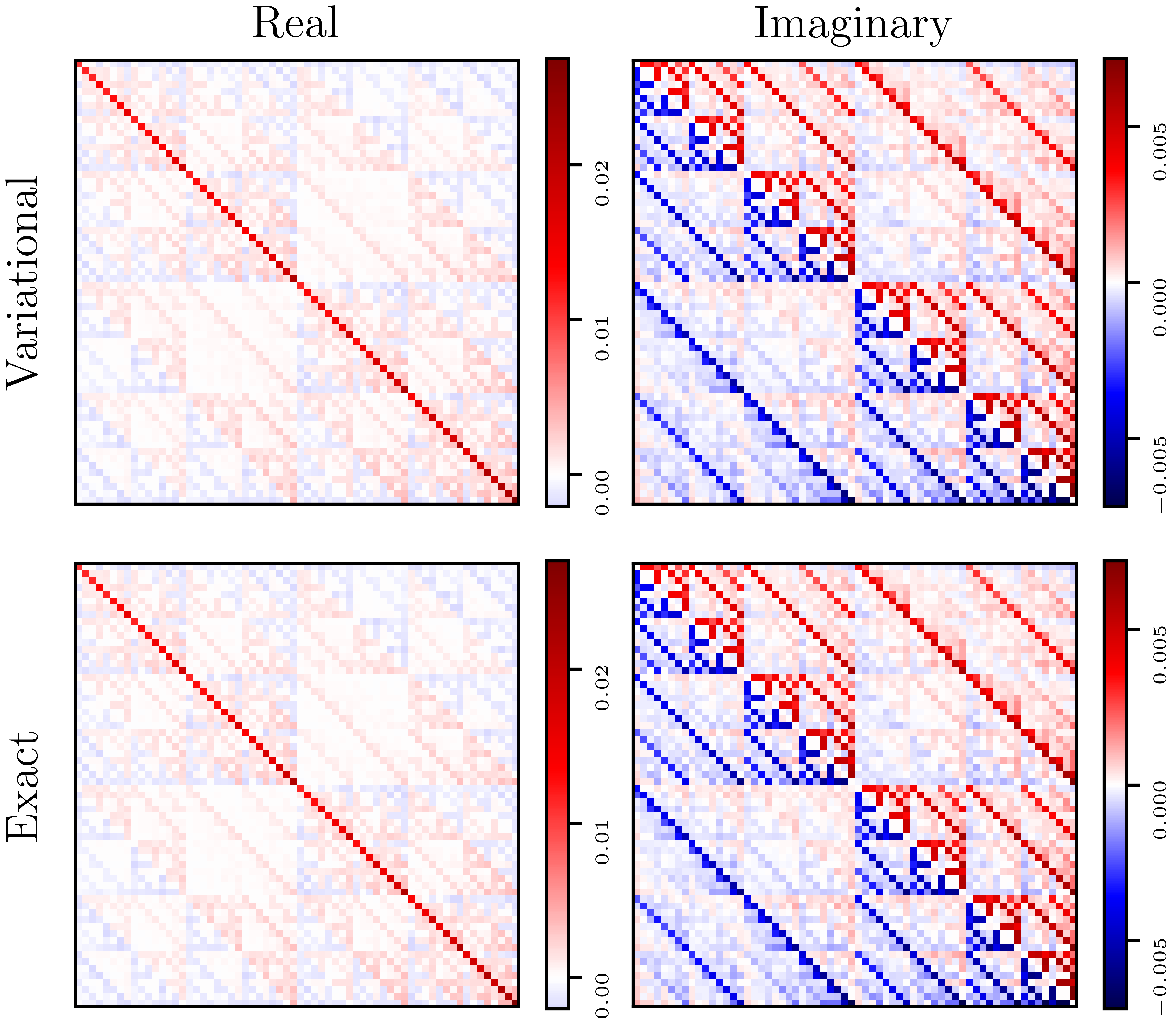}
    \caption{The real (left column) and imaginary (right column) parts of the variational (top row) and exact (bottom row) steady state density matrices of the DQIM with $N=6$ at $h=1.5$. The former is obtained after 1000 SR iterations, starting from a random initial MPO with $\chi=6$; the latter is obtained via exact diagonalization of the Lindbladian superoperator. The Uhlmann fidelity between the variational and exact density matrices is in excess of 0.99997.}
    \label{fig:density matrix}
\end{figure}

In Fig. \ref{fig:Bond dimension comparison} we investigate the convergence of VMPOMC as a function of increasing bond dimension $\chi$ for the DQIM with $N=12$ and $N=100$ spins. We concentrate on the cost function, steady-state magnetizations, and stead-state Rényi-2 entropy $S_2 = -\log_2 (Z)/N$. 
For $N=12$, we start the optimization process from entirely random initial tensors, while for $N=100$ we do so from the already converged, rescaled tensors obtained with $N=12$.
Expectedly, the final averaged values of the cost function decrease monotonically with increasing $\chi$.
The criterion for convergence ($\pazocal{C}/N<10^{-4}$) is satisfied for both system sizes for a bond dimension as low as $\chi=6$.
Correspondingly, respectable accuracy ($<4\%$ relative error) across all considered observables is achieved for $\chi=6$, and is improved further for $\chi\geq 10$, with $\approx 0.1\%$ relative error for $\chi=16$.
Most of the recorded measurements are visually indistinguishable between the two system sizes, which is indicative of the lack of discernible finite size effects when increasing the chain length from $12$ to $100$ spins for the chosen parameter values.
The low bond dimension required for convergence in our variational approach is worth contrasting with time-dependent MPO-based methods,
where the required bond dimensions may reach several hundred \cite{Cui}.

In Fig. \ref{fig:density matrix} we compare directly the variational density matrix after 1000 SR iterations to the exact steady state density matrix for a system of 6 spins. The variational density matrix approximates the steady state accurately, with an Uhlmann fidelity between the variational and exact density matrices in excess of 0.99997.

\begin{figure}[t]
    \includegraphics[width=\columnwidth]{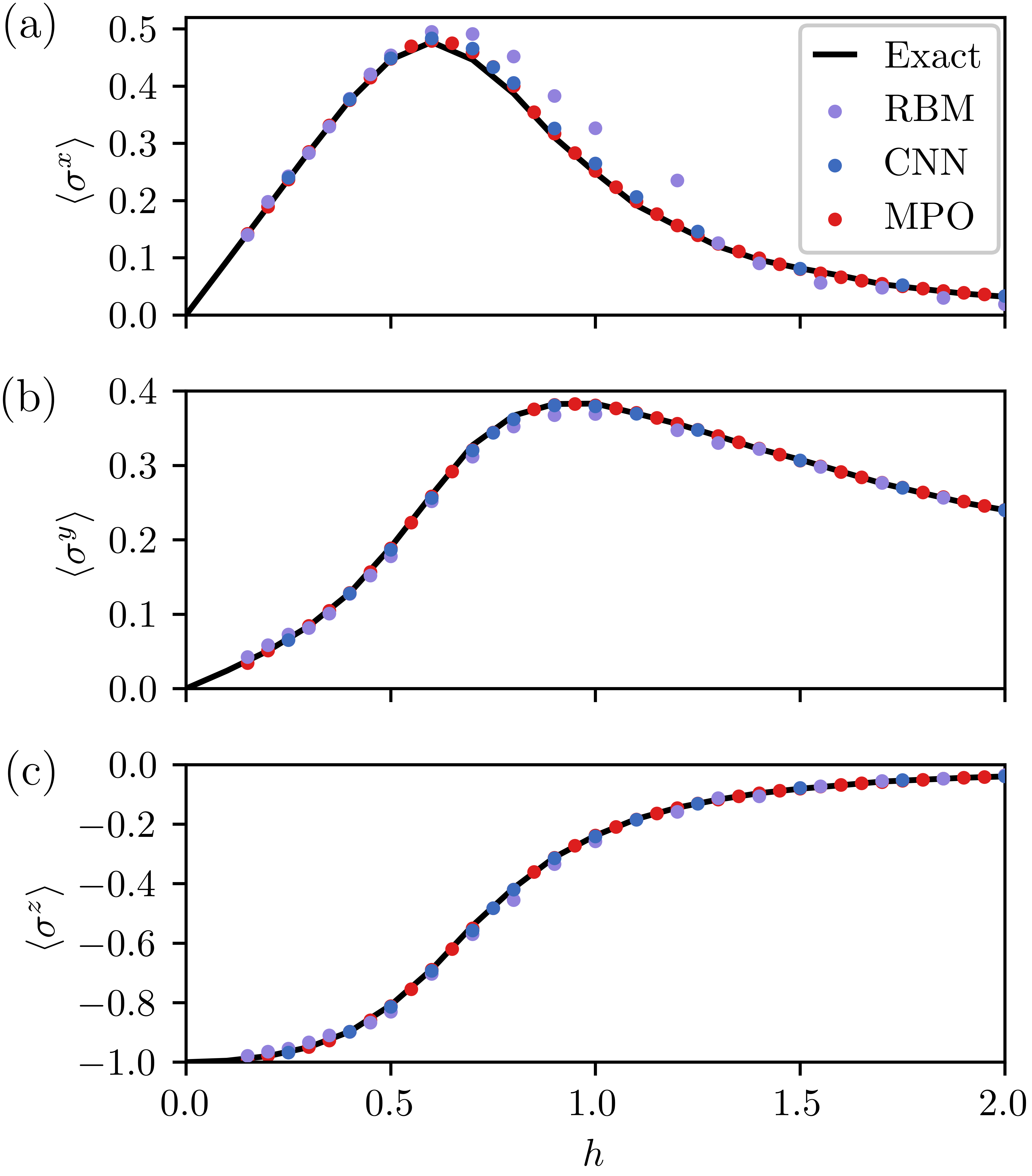}
    \caption{Steady-state magnetization phase diagrams for the DQIM with $N=16$. Our variational MPO-based results (red), obtained with $\chi=6$, are compared with variational RBM-based (purple) and exact (black) results from \cite{Vicentini2019}, as well as with variational CNN-based (blue) results from \cite{mellak2024deep}.}
    \label{fig:vicentini comparison}
\end{figure}

In Fig. \ref{fig:vicentini comparison} we compare the accuracy of our method against other time-independent variational approaches of \cite{Vicentini2019} and \cite{mellak2024deep}, based on restricted Boltzmann machine (RBM) and convolutional neural network (CNN) ansätze, respectively. Our steady-state magnetization results, obtained with an MPO of bond dimension $\chi=6$, are in the best agreement with the exact (quantum trajectories) results. 
We note that the total number of variational parameters used in our approach (144 for $\chi=6$) is close to 20 times smaller compared to the RBM ansatz (up to 2752), or 3 times smaller compared to the CNN ansatz (438). 
We attribute the improved performance of our method in part to the translationally-symmetric structure of the MPO ansatz, which reduces the number of variational parameters, as well as to the lack of additional statistical errors in the observable measurements.

\begin{figure}
    \includegraphics[width=\columnwidth]{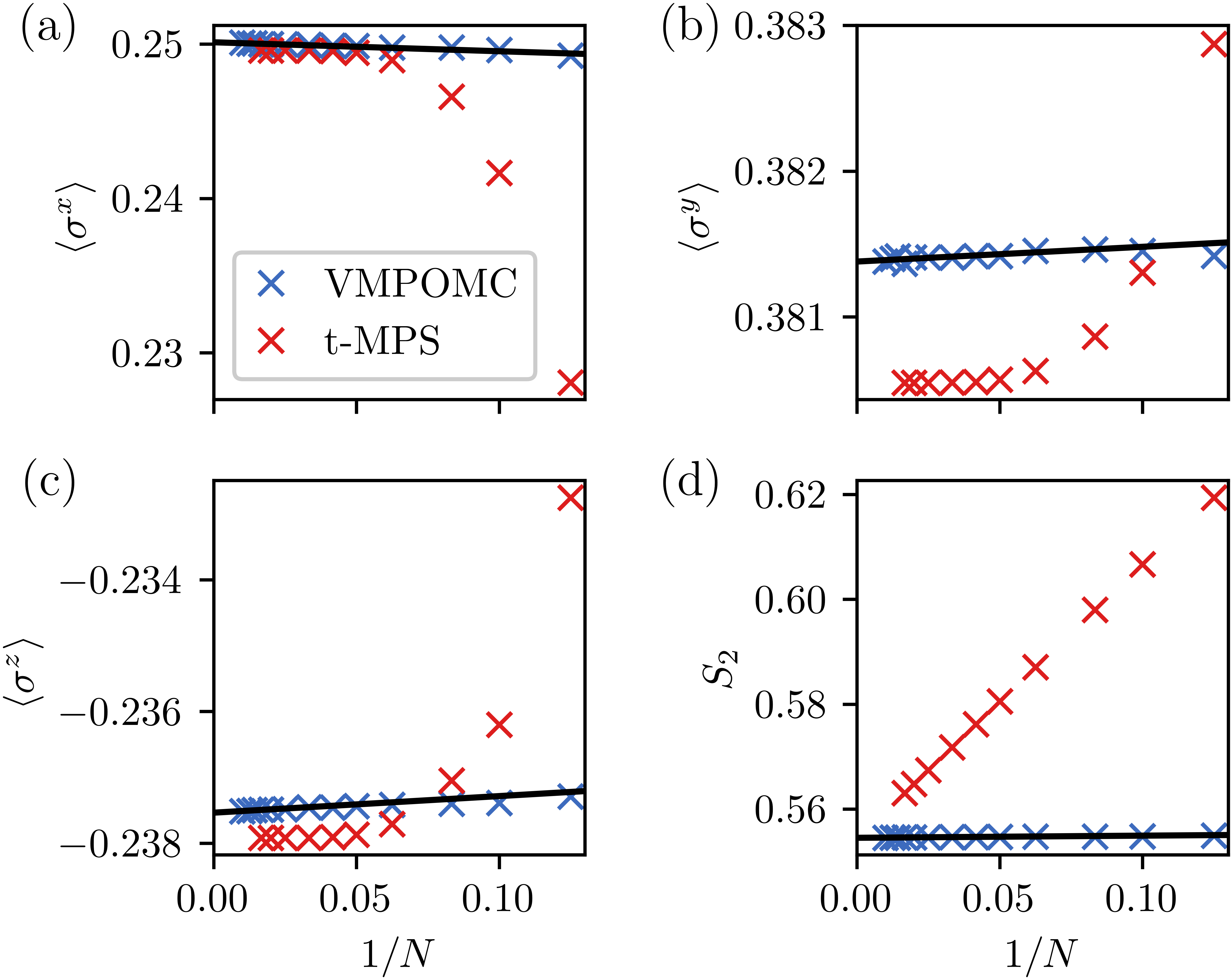}
    \caption{(a-c) Steady-state magnetizations and (d) Rényi-2 entropy for the dissipative quantum Ising chain with $\gamma_- = h = 2J = 1.0$ as a function of the inverse system size with $N$ between 8 and 100, obtained via VMPOMC for PBC and t-MPS for OBC with $\chi=10$ throughout. The solid lines are extrapolations of the VMPOMC results to the thermodynamic limit.}
    \label{fig: fse Ising}
\end{figure}

In Fig. \ref{fig: fse Ising} we perform a finite-size extrapolation of the steady-state magnetizations and Rényi-2 entropy, comparing the VMPOMC for the dissipative quantum Ising model with periodic boundary conditions (PBC) with a time-dependent MPS (or t-MPS) approach (as described in \cite{SCHOLLWOCK201196}) for an equivalent model with open boundary conditions (OBC), with the Hamiltonian
\begin{equation}
    H = J\sum_{i=1}^{N-1} \sigma^z_i \sigma^z_{i+1} + \frac{J}{2}(\sigma^z_1+\sigma^z_N) + h\sum_i \sigma^x_i.
\end{equation}
The steady-state expectation values are seen to converge to the thermodynamic limit with smaller system sizes for PBC than OBC.
In the latter case, a more accurate estimate of the thermodynamic expectation values is given by measuring the magnetizations of a single spin in the bulk of the chain instead of an average over the whole chain. On similar grounds, the estimate of $S_2$, a global property of the many-body system, converges much more slowly with increasing system size to its thermodynamic-limit value with OBC than with PBC. We attribute the slight deviations in the extrapolated thermodynamic limit-values for t-MPS from VMPOMC to additional sources of errors, such as due to the Trotter decomposition and finite time-step size.

\begin{figure}[t]
    \includegraphics[width=\columnwidth]{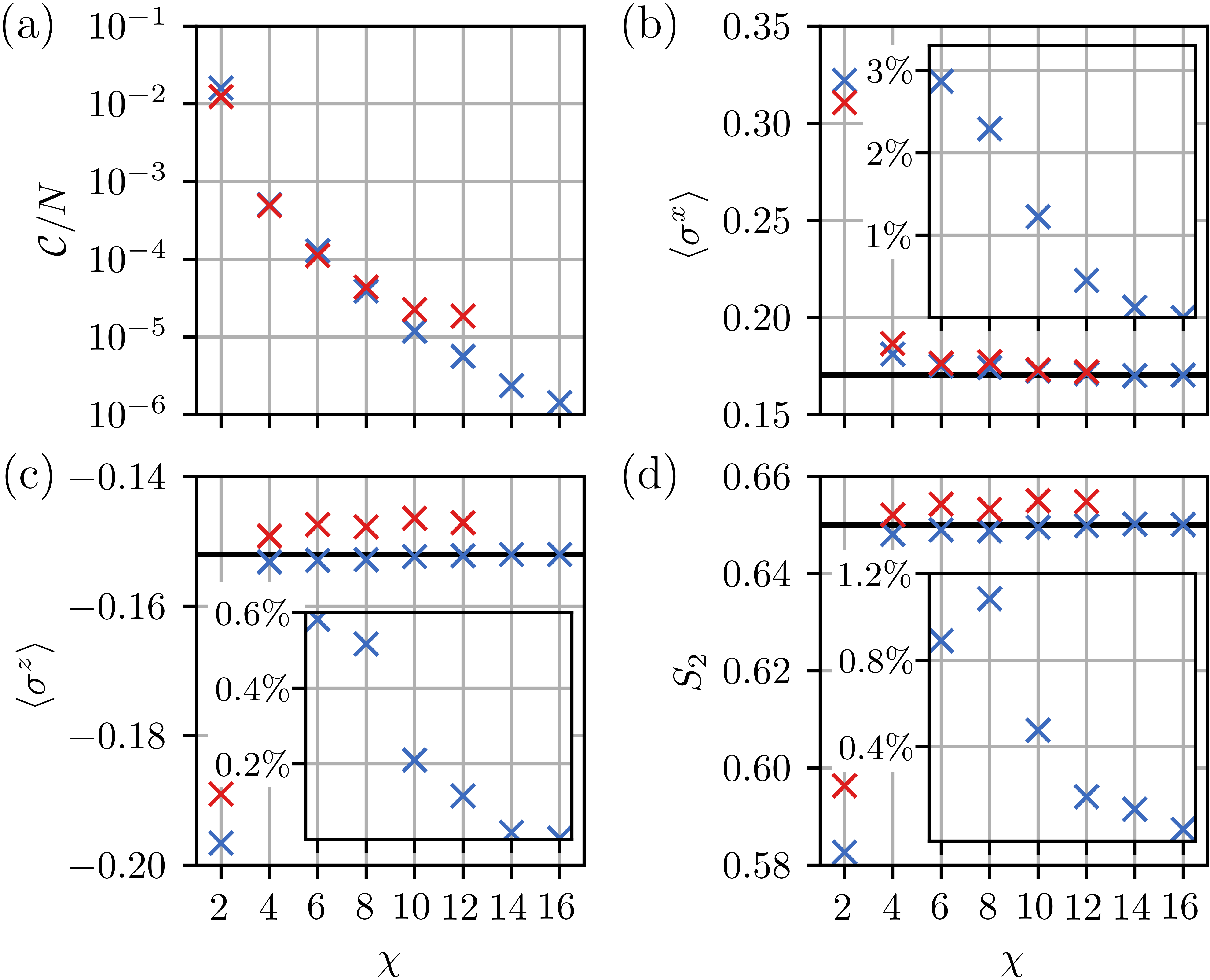}
    \caption{Convergence of VMPOMC for the DQIM with long-ranged interactions with $\alpha=2$ for $N=12$ (blue crosses) and $N=100$ (red crosses) at $h=1.0$ as function of the bond dimension $\chi$, showing the final recorded values of (a) the cost function, (b) steady-state $x$- and (c) $z$-magnetizations, and (d) 
    Rényi-2 entropy.
    Black horizontal lines in (b)-(d) indicate corresponding exact steady-state expectation values for $N=12$. Insets: relative errors between the variational and exact results for $N=12$ and $\chi\geq 6$ as function of $\chi$. }
    \label{fig:LRIsing bond dimension comparison}
\end{figure}

\subsection{Long-ranged interactions}

As a second example, we study the DQIM with spin decay as before, except with the nearest-neighbor interactions replaced by long-ranged ones. The corresponding Hamiltonian is
\begin{equation}
    H = \frac{J}{\pazocal{N}_\alpha}\sum_{j>i} \frac{\sigma^z_i \sigma^z_{j}}{|i-j|^\alpha} + h\sum_i \sigma^x_i, \label{long-ranged Ising Hamiltonian}
\end{equation}
where the exponent $\alpha \in (0,\infty)$ sets the decay length of the long-ranged interactions, and where $\pazocal{N}_\alpha$ is the Kac normalization factor \cite{Passarelli2022}.
These power law interactions are experimentally realizable via e.g. Rydberg atoms ($\alpha=3,6$) or trapped ions ($0<\alpha<3$) \cite{RevModPhys.95.035002}.

For $\alpha=\infty$ we recover the previously studied model with nearest-neighbor interactions. In the other limiting case of $\alpha=0$ the spins become fully connected; 
a family of such models has been studied extensively via a variety of mean-field \cite{PhysRevB.108.054302}, exact \cite{PhysRevLett.131.190403}, field-theoretic \cite{PhysRevA.104.023713}, semiclassical \cite{zhihao2023failure}, and other approaches \cite{Shammah2018,PhysRevB.102.094430,huybrechts2024quantum}.
For $0<\alpha<\infty$, variations on the above model have been studied only very recently with tree tensor network \cite{PhysRevA.109.022420} and truncated Wigner approaches \cite{PhysRevA.105.013716}.
Our MPO-based method allows for the exact treatment of long-ranged Ising interactions as in Eq. \eqref{long-ranged Ising Hamiltonian}, of, in principle, arbitrary decay lengths (contingent on sufficient expressibility of the MPO ansatz). This is made possible due to the purely diagonal form of the interaction Hamiltonian, which implies a cancellation of the probability amplitudes in Eqs. \eqref{local estimator of lindbladian} and \eqref{local derivative of lindbladian}, and thus absence of costly tensor contractions.

\begin{figure}[t]
    \includegraphics[width=\columnwidth]{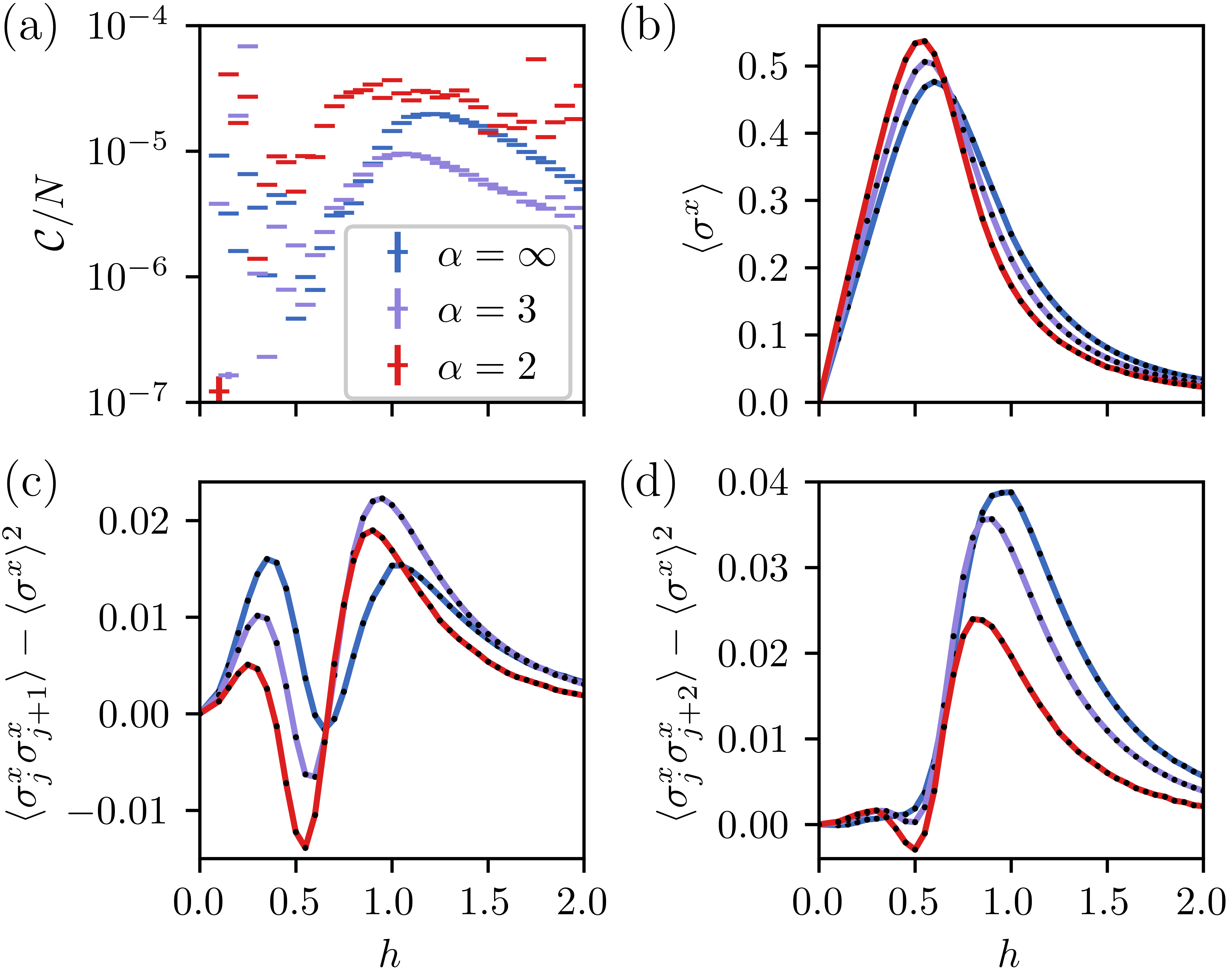}
    \caption{(a) Cost function, (b) steady-state magnetization, and (c) steady-state nearest-neighbor and (d) next nearest-neighbor spin-spin correlation functions for the DQIM with long-ranged interactions with $\alpha=2,3,\infty$ for $N=100$ and $\chi=10$, as function of the transverse-field strength $h$.}
    \label{fig:LRIsing phase diagrams}
\end{figure} 

The long-range interactions may induce long-ranged correlations whose simulation may be prohibitively difficult for the MPO ansatz.
We therefore begin by again assessing the convergence of our method with bond dimension for chains of length $N=12$ and $N=100$ at $h=1.0$ and $\alpha=2$ in Fig. \ref{fig:LRIsing bond dimension comparison}. 
As before, for $N=100$, we start from the already converged, rescaled tensors obtained with $N=12$.
Remarkably, convergence in both cases is achieved for modest bond dimensions of $\chi\geq 8$.

In Fig. \ref{fig:LRIsing phase diagrams} we study the effect of the long-ranged interactions on the magnetization phase diagram and spin-spin correlation functions of the dissipative quantum Ising model. We consider the decay exponents $\alpha=2,3,\infty$ with $N=100$ and $\chi=10$ throughout.
Remarkably, convergence was achieved for all considered decay exponents and external field strengths. 

\begin{figure}
    \includegraphics[width=\columnwidth]{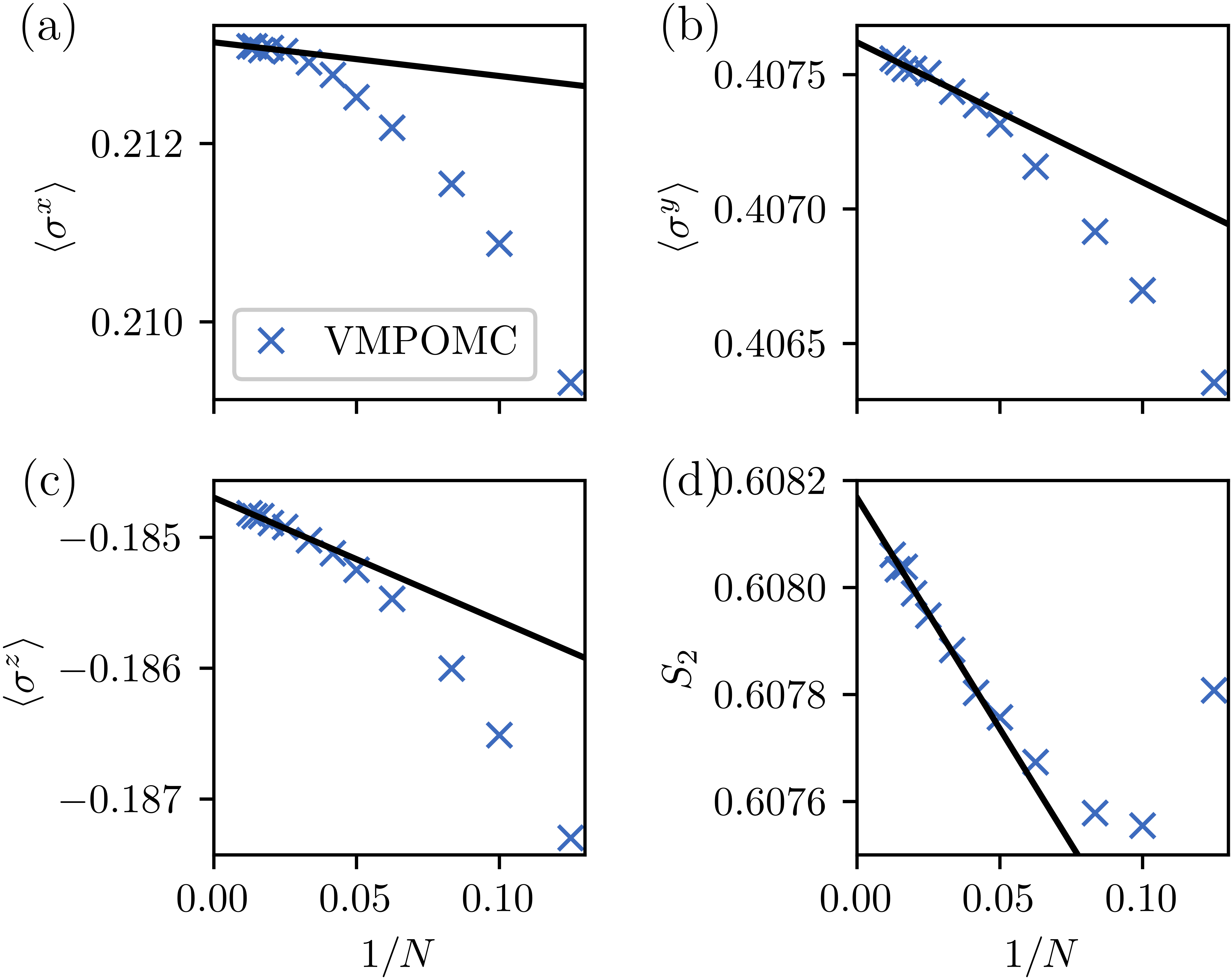}
    \caption{(a-c) Steady-state magnetizations and (d) Rényi-2 entropy for the dissipative quantum Ising chain with $\gamma_- = h = 2J = 1.0$ and long-ranged dipolar ($\alpha=3$) interactions as a function of the inverse system size with $N$ between 8 and 100, obtained via VMPOMC for PBC with $\chi=10$ throughout. The solid lines are extrapolations of the VMPOMC results to the thermodynamic limit.}
    \label{fig: fse LRIsing}
\end{figure}

In Fig. \ref{fig: fse LRIsing} we perform a finite-size extrapolation of the steady-state magnetizations and Rényi-2 entropy with the VMPOMC algorithm for the dissipative quantum Ising model with PBC and long-range dipolar interactions.

\subsection{Local and collective dephasing}

As a final example, we study the DQIM with nearest-neighbor interactions and single-site decay, alongside with a dephasing term. We consider two possible dephasing channels: local, described by the single-site Lindblad jump operators
$
    \Gamma_k^\text{(loc)} = \sqrt{\gamma_z}\sigma_k^z,
    \label{Lindblad operator dephasing local}
$
for all $k=1,\dots,N$,
and collective, described by the single non-local operator
$
    \Gamma^\text{(col)} = \sqrt{\gamma_z}\sum_{j=1}^N \sigma_j^z.
    \label{Lindblad operator dephasing collective}
$
Models with collective dephasing have attracted interest due to promising applications in e.g. quantum error prevention \cite{Kempe2001,Lidar2003}.
To the authors' best knowledge, the above DQIM with collective dephasing has not been studied in the present literature, with previous numerical and mean-field studies limited only to permutationally-invariant versions \cite{Shammah2018,PhysRevB.101.214302}. Similarly to the case of long-ranged Ising interactions, the exact treatment of this model with our method is made possible due to the purely-diagonal form of the Lindblad collective dephasing operator.

\begin{figure}
    \centering
    \includegraphics[width=\columnwidth]{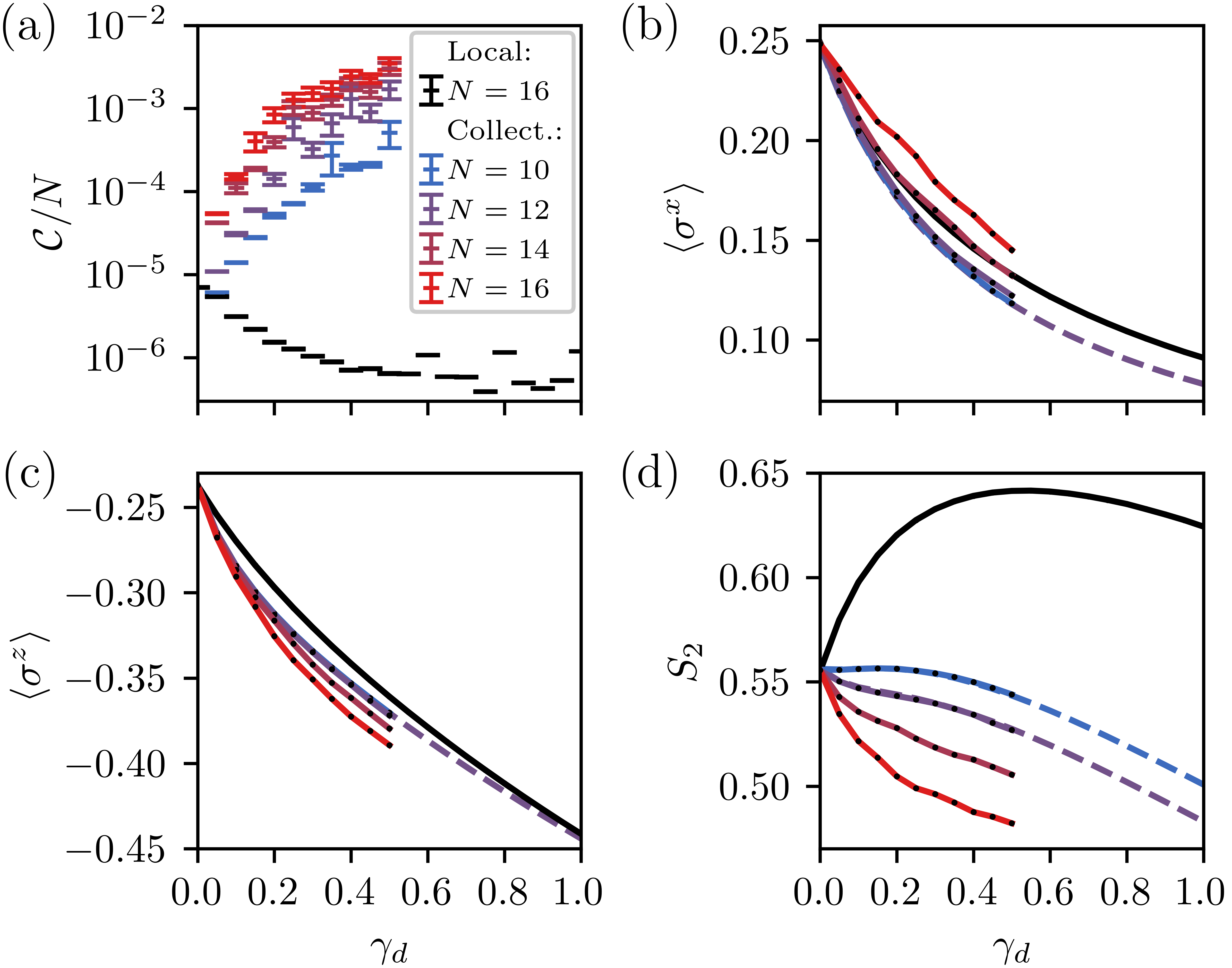}
    \caption{(a) Cost function, (b-c) steady-state magnetizations, and (d) steady-state Rényi-2 entropy for the DQIM with purely-local (black, $\chi=12$) and purely-collective (color, $\chi=24$) dephasing for up to $N=16$ at $h=1.0$ as function of the dephasing strength $\gamma_d$ (solid lines). Dashed lines in (b)-(d) represent exact results for up to $N=12$.}
    \label{fig:dephasing}
\end{figure}

In Fig. \ref{fig:dephasing} we compare the steady-state magnetization and Rényi-2 entropy $S_2$ 
phase diagrams in the presence of purely-local or purely-collective dephasing. Notably, we observe a striking difference in the behavior of $S_2$.
For local dephasing, $S_2$ can be seen to increase at weak dephasing before decreasing for $\gamma_d > 0.5$, whereas for collective dephasing the entropy decreases almost monotonically for all dephasing strengths. 
This behavior can be understood in terms of the rapid initial reduction of the off-diagonal elements (coherences) of the density matrix, resulting in a decrease in the purity, or an increase in $S_2$. This, however, also limits the coherent dynamics due to the external field counteracting spin decay in setting the populations, leading to the pure fully-decayed spin product state in the strong dephasing limit. For the model with collective dephasing, however, a subset of the coherences (of the decoherence-free subspace) survives, resulting in the near-immediate reduction in $S_2$.

Although the added computational overhead from both dephasing terms in computing the local estimator is negligible, the long-ranged correlations induced in the collective case prove too difficult to accurately describe with an MPO ansatz for larger dephasing strengths and system sizes, as indicated by higher recorded values of the cost function and poorer precision of the measured magnetizations and entropies in Fig. \ref{fig:dephasing}, despite using a twice larger bond dimension of $\chi=24$.
From a technical perspective, the collective dephasing term can be viewed as a dissipative analogue of an all-to-all spin-interaction, and hence is not expected to be simulable with a MPO ansatz.
More accurate simulation of such models may therefore be possible with e.g. long-ranged RBMs, which were shown to satisfy a volume-law entanglement entropy scaling for closed quantum systems \cite{PhysRevX.7.021021}.

\section{Conclusions and perspectives} \label{sec: discussion}

In this work, we introduced a novel time-independent variational Monte Carlo method, based on the optimization of a MPO tensor network ansatz, applicable to the non-equilibrium steady state of strongly-correlated open many-body quantum systems.
Our implementation for open quantum spin chains enjoys excellent agreement with exact results and superior accuracy over related variational Monte Carlo approaches at significantly smaller fraction of the number of variational parameters.
We delineated the unique advantages of VMPOMC over comparable approaches and illustrated its capabilities by simulating the one-dimensional dissipative quantum Ising model with highly non-local interactions, including collective dephasing and long-ranged power law interactions.
In particular, we have shown that the steady state can often be well described with an MPO of modest bond dimension, even in the presence of long-ranged interactions with as many as $N=100$ spins.

The main bottleneck preventing effective simulation of larger systems is the computation of the log-derivatives of the Lindbladian in Eq. \eqref{local derivative of lindbladian}, the cost of which scales as $N^2$ with the system size. In this regard, more advanced optimization techniques which further reduce the total required number of iterations, such as the linear method \cite{toulouse_optimization_2007}, may yield a significant improvement.

The proof-of-principle approach introduced in this work can be extended in several directions.
The translational invariance we imposed may be lifted, or the size of the unit cell may be increased,
albeit at increasing the number of variational parameters. Such an extension would enable the simulation of e.g. staggered systems with antiferromagnetic ordering in the steady state.
Non-equilibrium dynamics can be simulated by resorting to the time-dependent variational principle as was done in \cite{PhysRevLett.127.230501,Nagy2019,Hartmann} for quantum neural network ansätze.
Finally, it would be worthwhile to extend our method to two dimensions,
for instance by mapping the two-dimensional lattice onto an effective one-dimensional system with long-ranged interactions simulable with our approach, similarly to implementations of DMRG for closed two-dimensional systems \cite{doi:10.1146/annurev-conmatphys-020911-125018}.

\section*{Code availability}

The method introduced in this manuscript has been implemented in \textit{Julia} \cite{Bezanson2017}, and is accessible on GitHub \footnote{Online GitHub repository, \url{https://github.com/dhryniuk/VMPOMC}}.

\section*{Acknowledgements}

M. H. S. gratefully acknowledges financial support from Engineering and Physical Sciences Research Council (Grants No. EP/R04399X/1, No. EP/S019669/1, and No. EP/V026496/1).
This work was supported by the Engineering and Physical Sciences Research Council (Grant No. EP/R513143/1).
The authors acknowledge the use of the UCL HPC facilities and associated support services in the completion of this work.

\appendix

\section{Monte Carlo sampling} \label{sec: Monte Carlo sampling}

\subsection{Metropolis Monte Carlo} 

Many of the expectation values required by our method cannot be evaluated exactly due to the intractable size of the Hilbert space for larger numbers of sites. We have expressed these expectation values as ensemble averages with probability density $p(\bm{x})$, which we can approximate stochastically by sampling representative many-body configurations $\bm{x}$ from $p(\bm{x})$.
To do so, we resort to the Metropolis Markov chain Monte Carlo method, in which a Markov chain of states $\{\bm{x}_n\}_{n=1}^{N_\text{MC}}$ is constructed, whose stationary distribution approaches $p(\bm{x})$ in the limit $N_\text{MC}\rightarrow\infty$. Subsequent samples are accepted in accordance with the Metropolis probability,
\begin{equation} \label{Metropolis}
    p_\text{acc}(\bm{x}\rightarrow \bm{x}')
    = \min \left\{ 1,\frac{|\langle \bm{x}'|\rho \rangle|^2}{|\langle \bm{x}|\rho \rangle|^2} \right\}.
\end{equation} 
Although the simplest approach would be to draw new samples $\bm{x}'$ with a uniform proposal distribution, it is more efficient, given our MPO ansatz, to instead generate new samples sequentially, performing the Metropolis updates in Eq. \eqref{Metropolis} at each site of the chain in fixed order (detailed below). The new sample is recorded after a full sweep through chain is completed. 
As opposed to random Metropolis updates, the sequential Metropolis algorithm has a reduced overall computational cost of computing the probability amplitudes in Eq. \eqref{Metropolis} by making it possible to reuse previously computed left and right matrix products, analogously to the computation of the local estimator (see Appendix \ref{appendix Computation of the local estimator of the Lindbladian}).
Although this violates detailed balance, a condition normally required to converge towards the correct stationary distribution of the Markov chain, the Markov chain nevertheless converges to the correct stationary distribution, as shown in \cite{Manousiouthakis1999, Ren2006}. 

\subsection{Sequential Metropolis update} \label{appendix Sequential Metropolis sweep}

In this section we detail the rightward sequential Metropolis sweep, used to generate new samples $\bm{x}$ from the distribution $p(\bm{x})$ and new left matrix products $\{L_j\}_{j=1}^N$. A mirrored, leftward version of the below algorithm can be formulated very similarly. Note that to proceed with the algorithm, we must have access to a current sample $\bm{x}$ and set of right matrix products $\{R_j\}_{j=1}^N$, both known from a previous sweep; if this is the first sweep, we generate $\bm{x}$ at random and compute $\{R_j\}_{j=1}^N$ recursively (see Appendix \ref{appendix Computation of the local estimator of the Lindbladian}).

Compute $q=\tr R_N$. Then, for $j$ in $1, \dots, N$ in order:
\begin{enumerate}
    \item Draw a random new single-body spin configuration $x_j'$, different from $x_j$, from a uniform proposal distribution.
    \item Calculate the acceptance probability $p(x_j\leftarrow x_j') = q'/q$ where $q' = \tr L_{j-1} A(x_j') R_{N-j}$.
    \item Draw a random number $r$ from $\pazocal{U}[0,1]$. If $r<p(x_j\leftarrow x_j')$, set $x_j\leftarrow x_j'$ and $q\leftarrow q'$.
    \item Compute and store $L_j = L_{j-1}A(x_j)$.
\end{enumerate}
A single iteration of the above algorithm produces a new Monte Carlo sample $\bm{x}$ and a corresponding new set of left matrix products $\{L_j\}_{j=1}^N$, which can be used in a subsequent leftward sweep. The total computational cost is $3N\chi^3$. 

\subsection{Sample autocorrelation}

To explore the sample space efficiently, it is crucial that the Monte Carlo samples are independently distributed, i.e. subsequent samples are uncorrelated \cite{becca_sorella_2017}.
In addition to being efficiently calculable, the sequential Metropolis update produces uncorrelated Monte Carlo samples. To see this, we investigate the autocorrelation function, which for a Markov chain of states $\{\bm{x}_n\}_{n=1}^{N_\text{MC}}$ can be estimated via 
\begin{eqnarray}
    \Gamma_{\pazocal{O}}(t) = \frac{\sum_{n=1}^{N_\text{MC}-t}(\pazocal{O}(\bm{x}_{n})-\bar{\pazocal{O}})(\pazocal{O}(\bm{x}_{n+t})-\bar{\pazocal{O}})^*}{\sum_{n=1}^{N_\text{MC}-t}|\pazocal{O}(\bm{x}_{n})-\bar{\pazocal{O}}|^2}, 
\end{eqnarray}
where the integer $t$ denotes the separation between succesively accepted states of the Markov chain, and where $\pazocal{O}$ denotes an observable such that $\pazocal{O}(\bm{x}_n) = \bra{\bm{\beta}_n}\pazocal{O}\ket{\bm{\alpha}_n}$. We compute the mean value $\bar{\pazocal{O}}$ exactly from the MPO ansatz.

\begin{figure}[t]
    \centering
    \includegraphics*[width=\columnwidth]{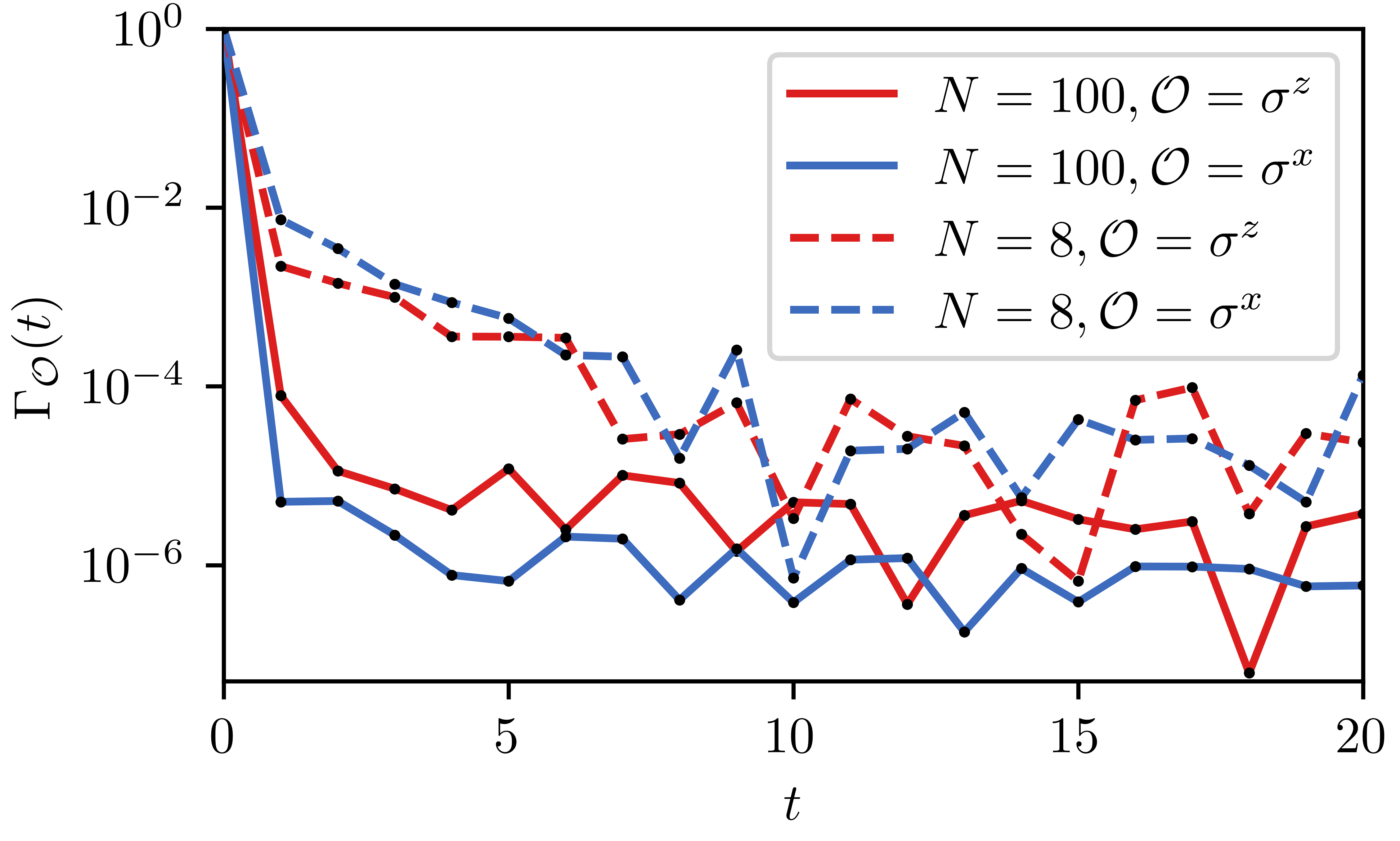}
    \caption{Autocorrelation functions between successive Monte Carlo samples for the DQIM for $N=8$ and $N=100$ at $h=0.3$, computed from converged MPO ansätze for $N_\text{MC}=10^6$ samples.}
    \label{fig: autocorrelation}
\end{figure}

In Fig. \ref{fig: autocorrelation} we plot example autocorrelation functions for Markov chains of Monte Carlo samples of length $N_\text{MC}=10^6$ for the DQIM of different system sizes from a converged steady-state MPO ansatz, and where the chosen observables are the spin-averaged magnetizations, e.g. $\pazocal{O}=\sigma^x = \frac{1}{N}\sum_{j=1}^N\sigma_j^x$. We draw a new Monte Carlo sample after a single left- or rightward sequential Metropolis sweep (in alternating directions), after an initial burn-in period of 1000 samples. For both system sizes the autocorrelations are already less than 0.01 after just a single sweep, and decrease even further for larger numbers of sweeps. For $N=100$, the recorded autocorrelations are on average smaller than for $N=8$, owing to the larger number of Metropolis spin flips in a single sequential update. Due to the low sample autocorrelation of the sequential Metropolis update, we deemed it sufficient to accept new Monte Carlo samples after only a single sweep for the purpose of the simulations discussed in the main text. 

\section{Efficient computation of the local estimator of the Lindbladian} \label{appendix Computation of the local estimator of the Lindbladian}

In this section we describe an efficient procedure to calculate the local estimator of the Lindbladian, in which we endeavor to minimize costly matrix multiplications. We shall specialize our discussion to the case of at most 2-local interactions allowed by the Lindbladian superoperator. 

To begin, it is helpful to split the Lindbladian into a 1-local part and a 2-local part,
\begin{equation}
    \pazocal{L} = \pazocal{L}^{(1)} + \pazocal{L}^{(2)},
\end{equation}
where $\pazocal{L}^{(i)}$ contains only $i$-local terms. 
Similarly, the local estimator of the Lindbladian decomposes as
\begin{equation}
    \pazocal{L}_\text{loc}(\bm{x}) = \pazocal{L}^{(1)}_\text{loc}(\bm{x}) + \pazocal{L}^{(2)}_\text{loc}(\bm{x}),
\end{equation}
which allows us to treat the two cases separately.
We shall now examine the 1-local case in detail.

The 1-local part of the Lindbladian is separable and can therefore be written as
\begin{equation}
    \pazocal{L}^{(1)} = \sum_{j=1}^N l^{(1)}_j \quad \text {where} \quad l^{(1)}_j = \mathbb{1}_{(1)}\otimes \dots \otimes l^{(1)}_{(j)} \otimes \dots \otimes  \mathbb{1}_{(N)},
\end{equation}
i.e. $\pazocal{L}^{(1)}$ can be expressed as a sum of tensor products $l^{(1)}_j$ of identities, with the exception of a single 1-body Lindbladian $l^{(1)}$ as the $j$-th term in the product.
We find, for the $1$-local part of the local estimator,
\begin{align*}
    \pazocal{L}^{(1)}_\text{loc}(\bm{x}) 
    &= \frac{1}{{\braket{\bm{x}}{\rho}}}\sum_{j=1}^N \sum_{\bm{y}} \left[\prod_{i=1}^N \bra{x_i}\left(l^{(1)}_j\right)_i\ket{y_i} \tr\prod_{i=1}^N A(y_i)\right] .\numberthis \label{loc1_derv}
\end{align*}
Observe that $\bra{x_i}(l^{(1)}_j)_i\ket{y_i} = \bra{x_j}l^{(1)}\ket{y_j}$ whenever $i=j$ and $\bra{x_i}(l^{(1)}_j)_i\ket{y_i} = \bra{x_i}\mathbb{1}\ket{y_i} = 1$ otherwise. In the latter case we have $y_i=x_i$ by orthogonality of the basis states, which also implies a cancellation of the probability amplitudes in Eq. \eqref{loc1_derv}. Hence, the above expression simplifies to
\begin{align}
    \pazocal{L}^{(1)}_\text{loc}(\bm{x}) &= \frac{1}{{\braket{\bm{x}}{\rho}}}\sum_{j=1}^N \sum_{\{y_j\}} \left[\bra{x_j}l^{(1)}\ket{y_j} \tr L_{j-1} A(y_j) R_{N-j} \right], \label{loc1}
\end{align}
where $L_{j-1}$ and $R_{N-j}$, defined via $L_j=\prod_{i=1}^{j}A(x_i)$ and $R_{j}=\prod_{i=N+1-j}^{N}A(x_i)$,
are the products of the matrices left and right of the matrix $A(y_j)$, respectively. Note the recurrence relations $L_j=L_{j-1}A(x_j)$ and $R_j=A(x_{N+1-j})R_{j-1}$ with $L_0=R_0=\mathbb{1}$. 

To see how the local estimator in Eq. \eqref{loc1} can be calculated efficiently, by minimizing the total number of required matrix multiplications, consider the following protocol of sequential updates of the $L_j$ and $R_j$ partial matrix products, given an initial $R_{N-1} = \prod_{i=2}^{N}A(x_i)$:
\begin{align*}
    &\pazocal{L}^{(1)}_\text{loc}(\bm{x}) \braket{\bm{x}}{\rho} = \sum_{\{y_1\}} \left[\bra{x_1}l^{(1)}\ket{y_1} \tr (A(y_1) R_{N-1}) \right] \\
    &+\sum_{\{y_2\}} \left[\bra{x_2}l^{(1)}\ket{y_2} \tr (\underbrace{A(x_1)}_{=L_1}A(y_2) \underbrace{A^{-1}(x_2)R_{N-1}}_{=R_{N-2}}) \right] \\
    &+\sum_{\{y_3\}} \left[\bra{x_3}l^{(1)}\ket{y_3} \tr (\underbrace{L_1A(x_2)}_{=L_2}A(y_3) \underbrace{A^{-1}(x_3)R_{N-2}}_{=R_{N-3}}) \right] \\
    &+\dots \numberthis
\end{align*}
The above procedure thus has a worst-case computational scaling of $\sim \!\!16N\chi^3$ compared to $\sim \!\!4N^2\chi^3$ in a naive approach. This can be reduced even further, to $\sim \!\!8N\chi^3$, if one already has access to the sets of partial matrix products $\{L_j\}_{j=1}^N$ and $\{R_j\}_{j=1}^N$. The latter is the approach used in our numerical implementation, where the sets $\{L_j\}_{j=1}^N$ and $\{R_j\}_{j=1}^N$ are constructed and memorised in advance, as part of prior left- or rightward sweeps. 

Another important simplification occurs for the diagonal elements of $l^{(1)}$, for which $y_j=x_j$, causing the probability amplitudes in Eq. \eqref{loc1} to cancel out. This simplification is valuable as it eliminates some of the tensor contractions entirely. For some operators, the one-body Lindbladian can in fact be dominated by diagonal elements, as is the case with e.g. the spin decay Lindblad jump operator, $\Gamma_k=\sigma_k^-$, whose corresponding $l^{(1)}$ matrix contains 4 non-zero elements, of which 3 lie on the diagonal.

The above discussion generalizes straightforwardly to the 2-local case, where we may write the Lindbladian as a sum of tensor products of identities and 2-body operators $l^{(2)}$, and where the sums over $y_j$ are replaced by sums over pairs of $y_j$ and $y_{j+1}$. The exception occurs at the boundary, where the operator $l_N^{(2)}$ connects the pair of spins labeled by $y_N$ and $y_{1}$ (which are nevertheless still 2-local due to periodic boundary conditions). In the same vein, we can generalize this procedure to arbitrary $n$-body or $n$-local interactions, albeit at a rapidly increasing size of the local $l^{(n)}$ matrices or the requirement for additional matrix multiplications in connecting non-local couplings.

We point out that comparably performant algorithms for the computation of the local estimator and log-derivatives of the Lindbladian may be also devised based on a direct contraction of the Lindbladian and differential operators with the MPO density matrix, without explicit decomposition over the $\{\bm{y}\}$ sub-states (first equalities in \eqref{local estimator of lindbladian} and \eqref{local derivative of lindbladian}), albeit at possibly higher memory costs and without enabling the discussed simplifications for non-local interactions arising from the diagonality of the Lindbladian operators.

\section{Efficient computation of the gradient} \label{appendix Computation of the gradient}

In this section we discuss how to efficiently compute the gradients in Eq. \eqref{SGD gradient}.
The main difference from computing the local estimator stems from the added presence of the log-derivative tensor $\Delta_{uv}^s (\bm{y})$. As such, we must first discuss how to differentiate a translationally-invariant MPO with respect to a tensor element. 
The probability amplitude for a translationally-invariant MPS is given by
\begin{equation}
    \bra{\bm{\alpha}}\ket{\psi} = \tr A(\alpha_1)A(\alpha_2)\dots A(\alpha_N),
\end{equation}
Differentiating with respect to the matrix element $a^s_{uv}$ of $A(s)$ one obtains
\begin{align*}
    \frac{\partial \bra{\bm{\alpha}}\ket{\psi}}{\partial a^s_{uv}} &= \frac{\partial }{\partial a^s_{uv}} \sum_k \sum_{\{(j)\}} a^{\alpha_1}_{k(1)} a^{\alpha_2}_{(1)(2)}\dots a^{\alpha_N}_{(N-1)k} \\
    &= \sum_k \sum_{\{(j)\}} [ \delta_{s\alpha_1} \delta_{uk} \delta_{v(1)} a^{\alpha_2}_{(1)(2)}\dots a^{\alpha_N}_{(N-1)k} \\&\qquad + \delta_{s\alpha_2} \delta_{u(1)} \delta_{v(2)} a^{\alpha_1}_{k(1)}\dots a^{\alpha_N}_{(N-1)k} + \dots ] \\
    &= \sum_{j=1}^N \delta_{s\alpha_j} \sum_k r_{vk}^{(N-j)} l_{ku}^{(j-1)}, \numberthis \label{derivative of gradient derivation}
\end{align*}
where the lower case $l$ and $r$ denote the matrix elements of the corresponding partial matrix products defined earlier, $L_j = [l^{(j)}_{uv}]$ and $R_j = [r^{(j)}_{uv}]$. 
This result is different to the one stated in Eq. (9) in \cite{Sandvik2007} and is found to lead to more stable convergence. The above carries over to a MPO when vectorized into a MPS.

Hence, to efficiently compute the log-derivative tensor $\Delta_{uv}^s (\bm{y})$ elements, 
we first compute the set of all left and right partial matrix products $\{L_j\}_{j=1}^{N}$ and $\{R_j\}_{j=1}^{N}$ for the state $\bm{y}$ recursively, from where Eq. \eqref{derivative of gradient derivation} yields
\begin{equation}
    \Delta_{uv}^s (\bm{y}) = \frac{1}{\bra{\bm{y}}\ket{\rho}}\sum_{j=1}^N \delta_{sx_j} \sum_k r_{vk}^{(N-j)} l_{ku}^{(j-1)}. \label{log derivative tensor full}
\end{equation}
The total computational cost amounts to $2N\chi^3$ from the computation of the left and right matrix products and a further $N\chi^3$ from the computation of the numerator in Eq. \eqref{log derivative tensor full}, for a total cost of $3N\chi^3$. The cost of computing the denominator is neglected as it's already known from computing the cost function earlier.

With $\Delta_{uv}^s (\bm{y})$ known, the computation of the ensemble averages in Eq. \eqref{SGD gradient} can again be performed in a sweeping manner, analogous to the computation of the cost function as described in Appendix \ref{appendix Computation of the local estimator of the Lindbladian}.

\section{Parameters and hyperparameters} \label{app: Parameters and hyperparameters}

In the table below we collected the parameter and hyperparameter values of all calculations discussed in the main text.
The column $N_\text{MC}$ lists the number of Monte Carlo samples per each Markov chain, per iteration. The column $N_\text{work.}$ lists the number of independent Markov chains per iteration (distributed over different MPI processes). The column $N_\text{iter.}$ lists the total number of iterations completed before the optimization process was terminated. In all simulations, the iteration step size $\delta$ decreases geometrically with the iteration number $k$ as $\delta=\delta_0 F^k$; the values $\delta_0$ and $F$ used are listed in the final two columns.

\onecolumngrid

\begin{center}
    \renewcommand{\tabcolsep}{3pt}
    \begin{tabular}{|c|c|c|c|c|c|c|c|c|c|c|c|c|c|c|c|} 
     \hline
     Figure & $N$ & $J$ & $h$ & $\gamma_l$ & $\gamma^\text{loc}_d$ & $\gamma^\text{col}_d$ & $\alpha$ & Method ($\epsilon$) & $\chi$ & $N_\text{MC}$& $N_\text{work.}$ & $N_\text{iter.}$ & $\delta_0$ & $F$ \\
     \hline 
     \ref{fig:SGD SR comparison} (blue) & 12 & 0.5 & 0.5 & 1.0 & 0 & 0 & --- & SGD & 4 & 160 & 6 & --- & 0.003 & 0.998 \\
     \ref{fig:SGD SR comparison} (red) & 12 & 0.5 & 0.5 & 1.0 & 0 & 0 & --- & SR (0.01) & 4 & 160 & 6 & --- & 0.03 & 0.998 \\
     \ref{fig:density matrix} & 6 & 0.5 & 1.5 & 1.0 & 0 & 0 & --- & SR (0.1) & 6 & 360 & 6 & 1000 & 0.01 & 0.998 \\
     \ref{fig:vicentini comparison} & 16 & 0.5 & \text{var.} & 1.0 & 0 & 0 & --- & SR (0.1) & 6 & 1000 & 6 & 10000 & 0.005 & 0.9995 \\
     \ref{fig: fse Ising} (blue) & \text{var.} & 0.5 & 1.0 & 1.0 & 0 & 0 & --- & SR (0.1) & 10 & 500 & 10 & 5000 & 0.02 & 0.9995 \\
     \ref{fig:Bond dimension comparison} (blue) & 12 & 0.5 & 1.0 & 1.0 & 0 & 0 & --- & SR (0.001) & \text{var.} & 400 & 40 & 50000 & 0.02 & 0.9999 \\
     \ref{fig:Bond dimension comparison} (red) & 100 & 0.5 & 1.0 & 1.0 & 0 & 0 & --- & SR (0.001) & \text{var.} & 400 & 160 & 50000 & 0.02 & 0.9999 \\
     \ref{fig:LRIsing bond dimension comparison} (blue) & 12 & 0.5 & 1.0 & 1.0 & 0 & 0 & 2 & SR (0.001) & \text{var.} & 400 & 40 & 50000 & 0.02 & 0.9999 \\
     \ref{fig:LRIsing bond dimension comparison} (red) & 100 & 0.5 & 1.0 & 1.0 & 0 & 0 & 2 & SR (0.001) & \text{var.} & 400 & 160 & 50000 & 0.02 & 0.9999 \\
     \ref{fig:LRIsing phase diagrams} (blue, $h < 0.3$) & 100 & 0.5 & \text{var.} & 1.0 & 0 & 0 & $\infty$ & SR (0.1) & 10 & 1000 & 16 & 3000 & 0.001 & 0.9995 \\
     \ref{fig:LRIsing phase diagrams} (blue, $h \geq 0.3$) & 100 & 0.5 & \text{var.} & 1.0 & 0 & 0 & $\infty$ & SR (0.1) & 10 & 250-500 & 16 & 10000 & 0.003 & 0.9997 \\
     \ref{fig:LRIsing phase diagrams} (purple, $h < 0.3$) & 100 & 0.5 & \text{var.} & 1.0 & 0 & 0 & 3 & SR (0.1) & 10 & 1000 & 16 & 3000 & 0.001 & 0.9995 \\
     \ref{fig:LRIsing phase diagrams} (purple, $h \geq 0.3$) & 100 & 0.5 & \text{var.} & 1.0 & 0 & 0 & 3 & SR (0.1) & 10 & 250-500 & 16 & 11000 & 0.003 & 0.9997 \\
     \ref{fig:LRIsing phase diagrams} (red, $h < 0.3$) & 100 & 0.5 & \text{var.} & 1.0 & 0 & 0 & 2 & SR (0.1) & 10 & 1000 & 16 & 3000 & 0.001 & 0.9995 \\
     \ref{fig:LRIsing phase diagrams} (red, $h \geq 0.3$) & 100 & 0.5 & \text{var.} & 1.0 & 0 & 0 & 2 & SR (1.0) & 10 & 250-500 & 16 & 13000 & 0.003 & 0.9998 \\
     \ref{fig: fse LRIsing} (blue) & \text{var.} & 0.5 & 1.0 & 1.0 & 0 & 0 & 3 & SR (0.1) & 10 & 500 & 10 & 5000 & 0.02 & 0.9995 \\
     \ref{fig:dephasing} (black) & 16 & 0.5 & 1.0 & 1.0 & \text{var.} & 0 & --- & SR (0.1) & 12 & 500 & 16 & 8000 & 0.005 & 0.9995 \\
     \ref{fig:dephasing} (colour) & \text{var.} & 0.5 & 1.0 & 1.0 & 0 & \text{var.} & --- & SR (0.1-1.0) & 24 & 500 & 16 & 12000 & 0.005 & 0.9996 \\
     \hline
    \end{tabular}
\end{center}

\twocolumngrid

\bibliography{export.bib}

\end{document}